\documentclass[twocolumn,showpacs,aps,prb]{revtex4-1}
\usepackage{graphicx}
\usepackage{epstopdf}
\usepackage{bm}% bold math
\usepackage{subfigure}
\usepackage{sidecap}

\usepackage{graphicx}
\usepackage{epstopdf}
\usepackage{amssymb}
\usepackage{amsmath}

\usepackage{bm}% bold math
\usepackage{subfigure}
\usepackage{sidecap}

\newcommand{\si}{\sigma}
\newcommand{\al}{\alpha}

\newcommand{\ga}{\gamma}

\newcommand{\ka}{\kappa}
\newcommand{\vare}{\varepsilon}
\newcommand{\de}{\delta}
\newcommand{\De}{\Delta}
\newcommand{\bsig}{{\bm\sigma}}

\newcommand{\bphi}{{\bm\phi}}

\newcommand{\hphi}{\hat{\phi}}

\newcommand{\be}{\begin{equation}}
\newcommand{\ee}{\end{equation}}

\newcommand{\bea}{\begin{eqnarray}}
\newcommand{\eea}{\end{eqnarray}}
\newcommand{\bd}{\begin{displaymath}}
\newcommand{\ed}{\end{displaymath}}
\newcommand{\ba}{\begin{array}}
\newcommand{\ea}{\end{array}}
\newcommand{\bi}{\begin{itemize}}
\newcommand{\ei}{\end{itemize}}
\newcommand{\bc}{\begin{center}}
\newcommand{\ec}{\end{center}}
\newcommand{\bfl}{\begin{flushleft}}
\newcommand{\efl}{\end{flushleft}}
\newcommand{\bfr}{\begin{flushright}}
\newcommand{\efr}{\end{flushright}}
\newcommand{\non}{\nonumber}

\newcommand{\bl}{\begin{aligned}}
\newcommand{\el}{\end{aligned}}

\newcommand{\hh}{\hat{h}}
\newcommand{\hG}{\hat{G}}

\newcommand{\ha}{\hat{a}}
\newcommand{\hb}{\hat{b}}
\newcommand{\hla}{\hat{\lambda}}

\newcommand{\oi}{{\rm i}}

\newcommand{\tA}{\tilde{A}}

\newcommand{\tilh}{\tilde{h}}

\newcommand{\tiD}{\tilde{D}}

\newcommand{\fa}{\frac{a}{2}}
\newcommand{\fc}{\frac{c}{2}}
\newcommand{\fl}{\frac{3}{2}}
\newcommand{\fu}{\frac{5}{2}}

\newcommand{\om}{\oi\omega_n}

\newcommand{\CC}{CeCoIn$_5$}

\newcommand{\URU}{UR\lowercase{u}$_2$S\lowercase{i}$_2$}

\def\ket#1{\left\vert #1 \right\rangle}
\def\dg{^{\dagger}}

\def\bR{{\bf R}}
\def\bk{{\bf k}} \def\bK{{\bf K}}\def\bq{{\bf q}} 
 
\def\bQ{{\bf Q}} 
  
 \def\bd{{\bf d}}  
  \def\hbz{\hat{{\bf z}}}
  
\def\bbk{\bar{\bk}}

\def\dg{\dagger}

\def\bra{\langle}
\def\ket{\rangle}
\def\={\!\!\!&=&\!\!\!}
\def\+{\!\!\!&&\!\!\!+~}
\def\-{\!\!\!&&\!\!\!-~}

\begin{document}
\date{\today}
\title{Hidden order symmetry and superconductivity in URu$_2$Si$_2$ investigated by quasiparticle interference}
%\author{N.N.}
\author{Alireza Akbari$^1$}
\author{Peter Thalmeier$^2$}
\affiliation{$^1$Max Planck Institute for Solid State Research, D-70569 Stuttgart, Germany\\
$^2$Max Planck Institute for the  Chemical Physics of Solids, D-01187 Dresden, Germany}

\begin{abstract}
The hidden order (HO) in \URU~ has been determined as a high rank multipole formed by
itinerant  5f-electrons with distinct orbital structure imposed by the crystalline electric field. Because this can
lead to a considerable number of different multipoles it is of great importance to use microscopic techniques
that are sensitive to their subtle physical differences. Here we investigate whether quasiparticle interference (QPI) method
can distinguish between the two most frequently proposed HO parameter models: the even rank-4 hexadecapole and the 
odd-rank-5 dotriacontapole model. We obtain the quasiparticle dispersion and reconstructed Fermi surface in each HO phase
adapting an effective two-orbital model of 5f bands that reproduces the main Fermi surface sheets of the para phase.
We show that the resulting QPI spectrum reflects directly the effect of fourfold symmetry breaking in the rank-5 model 
which is absent in the rank-4 model. Therefore we suggest that QPI method should give a possibility of direct discrimination 
between the two most investigated models of HO in \URU. Furthermore the signature of proposed chiral d-wave superconducting (SC) order parameter in QPI of the coexisting HO+SC phase is investigated. 
\end{abstract}


\maketitle

\section{Introduction}
The nature of hidden order (HO) in  \URU~ is considered a central topic in the investigation of strongly correlated electron systems \cite{mydosh:11}. For its theoretical analysis several fundamental issues arise. Initially the HO was described in terms of tetragonal crystalline electric field (CEF) split localized $5f^2$ (U$^{4+}$) states \cite{santini:94}. Their localized multipoles would then experience effective RKKY-type inter-site interactions leading to their long range order below T$_{HO}=17.5$ K. In fact thermodynamic properties of the HO transition may be described within the localized context \cite{santini:94}.

However later ARPES experiments \cite{meng:13} and theoretical analysis \cite{oppeneer:10} suggested that the 5f electrons have itinerant character and their Fermi surface (FS) reconstruction below $T_{HO}$ plays an essential role in the HO mechanism. The hidden order parameter then should be constructed from itinerant $5f$ basis states rather than localized ones. This was carried out by Ikeda et al \cite{ikeda:12} within an extended many-body model starting from band structure calculations. It was found that the antiferro-type HO evolves due to a nesting between $\Gamma$ and Z-centered electron and hole pockets with a wave vector $\bQ=(0,0,1)$. Because they are mainly formed by orbitals with large total angular momentum ($j=\frac{5}{2}$) component, $M=\pm\frac{3}{2},\pm\frac{5}{2}$,  the dominating AF hidden order parameter is a multipole of rank-5 $(E_-)$ type which breaks translational, c-axis $C_4$  rotational and time reversal symmetries. The evolution of this order leads to characteristic reconstruction of the Fermi surface: Due to doubling of the unit cell the $Z$-~point hole pocket is downfolded to the $\Gamma$-point of the Brillouin zone (BZ) and electron and hole pockets are broken into the smaller FS sheets at their crossing  points. Concurrently a HO gap evolves in the density of states (DOS).

Deep inside the HO phase unconventional heavy fermion superconductivity appears at T$_c=1.5$ K which was suggested to have chiral d-wave symmetry \cite{kasahara:07}, but this sofar remains a conjecture. It coexists homogeneously with HO and vanishes at the same critical pressure $p_c\simeq 0.75$ kbar.
 Recently the quasiparticle interference (QPI) method has been proven very succesful in unraveling the gap symmetry of heavy fermion superconductors \cite{akbari:11,allan:13,zhou:13}. Already before \cite{schmidt:10} the method was demonstrated in the HO phase of \URU~ but not yet in the coexistence region with superconductivity (SC+HO). 
 
Here we present a theoretical analysis of QPI both in the HO state and coexisting HO+SC phase. Our main goal is to understand the principal effects which the FS reconstruction due to HO has on the QPI and whether this holds any clue to the symmetry of the HO phase. For that purpose we make a comparative analysis of reconstructed bands and Fermi surfaces as well as QPI spectra for the most frequently involved HO symmetries, namely the doubly degenerate rank-5 E$_-$ dotriacontapole introduced above \cite{ikeda:12} and the non-degenerate antiferro-type rank-4 hexadecapole \cite{santini:94,haule:09}. The latter only breaks translational symmetry and tetragonal in-plane reflectional symmetry but preserves c-axis $C_4$ rotational and time reversal symmetry. 

For this purpose we start from an effective 5f electron band model describing the $\Gamma$ and Z-point electron and hole pockets (but not the small pockets on the A and M points of folded BZ corners) which was introduced by Rau and Kee \cite{rau:12}.  We give closed expressions for the reconstructed quasiparticle bands in the HO phase for both models in the whole BZ. Using this result we can calculate with high accuracy the expected QPI spectrum, map its characteristic structures and relate them to the reconstructed HO Fermi surface. We also discuss possible connections to the experimental results \cite{schmidt:10}. Finally we include a BCS term for the reconstructed HO bands with a SC gap symmetry of the chiral d-wave type. This allows us to predict the QPI spectrum in the coexisting HO+SC phase which has not yet been performed experimentally. We will finally discuss the features in QPI that may be taken as typical consequence of the chiral d-wave symmetry.

\section{Two-orbital model of heavy electron bands in \URU}

\begin{figure*}
\vspace{0.2cm}
\includegraphics[width=50mm,clip]{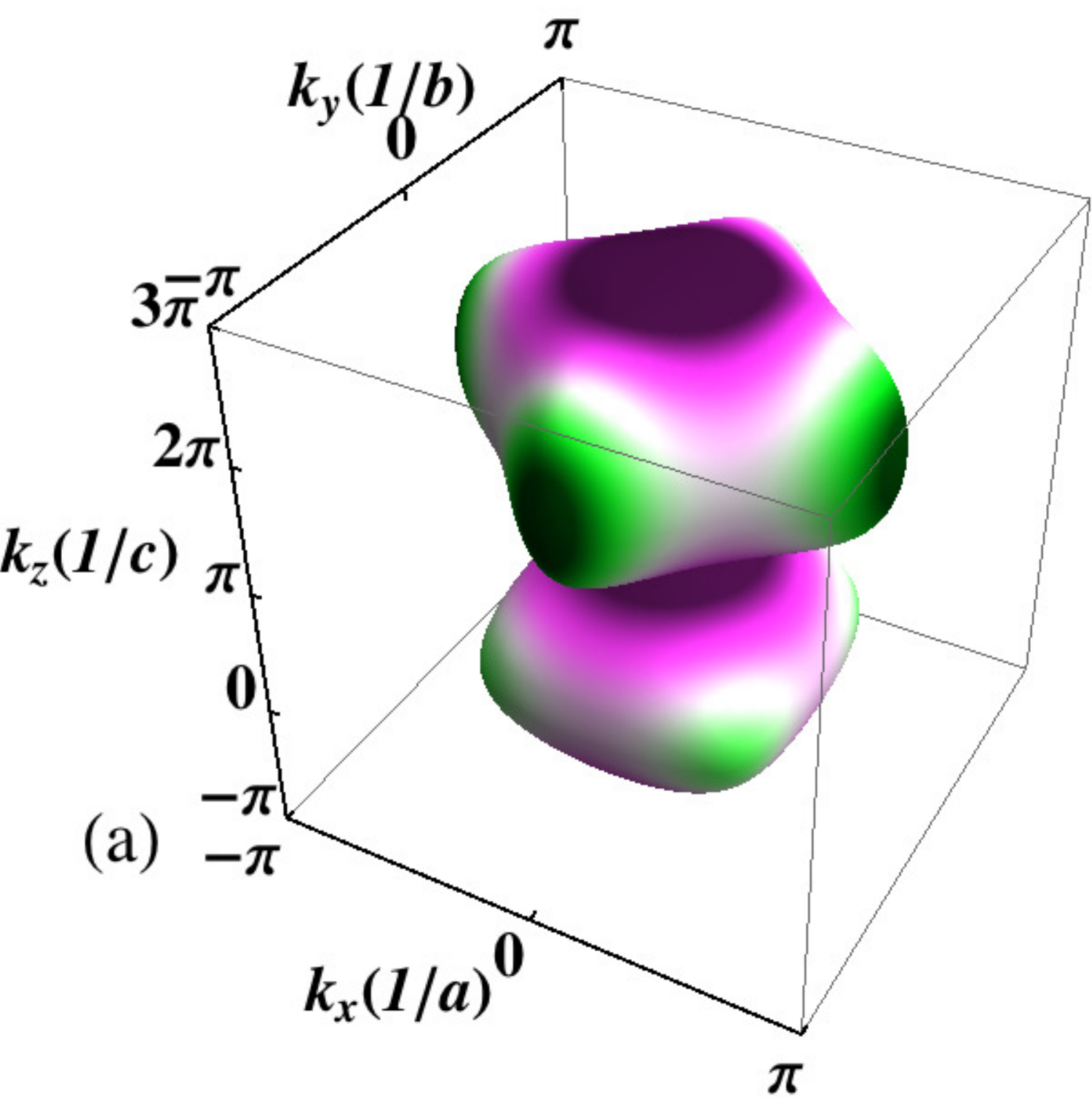}\hspace{0.3cm}
\includegraphics[width=95mm,clip]{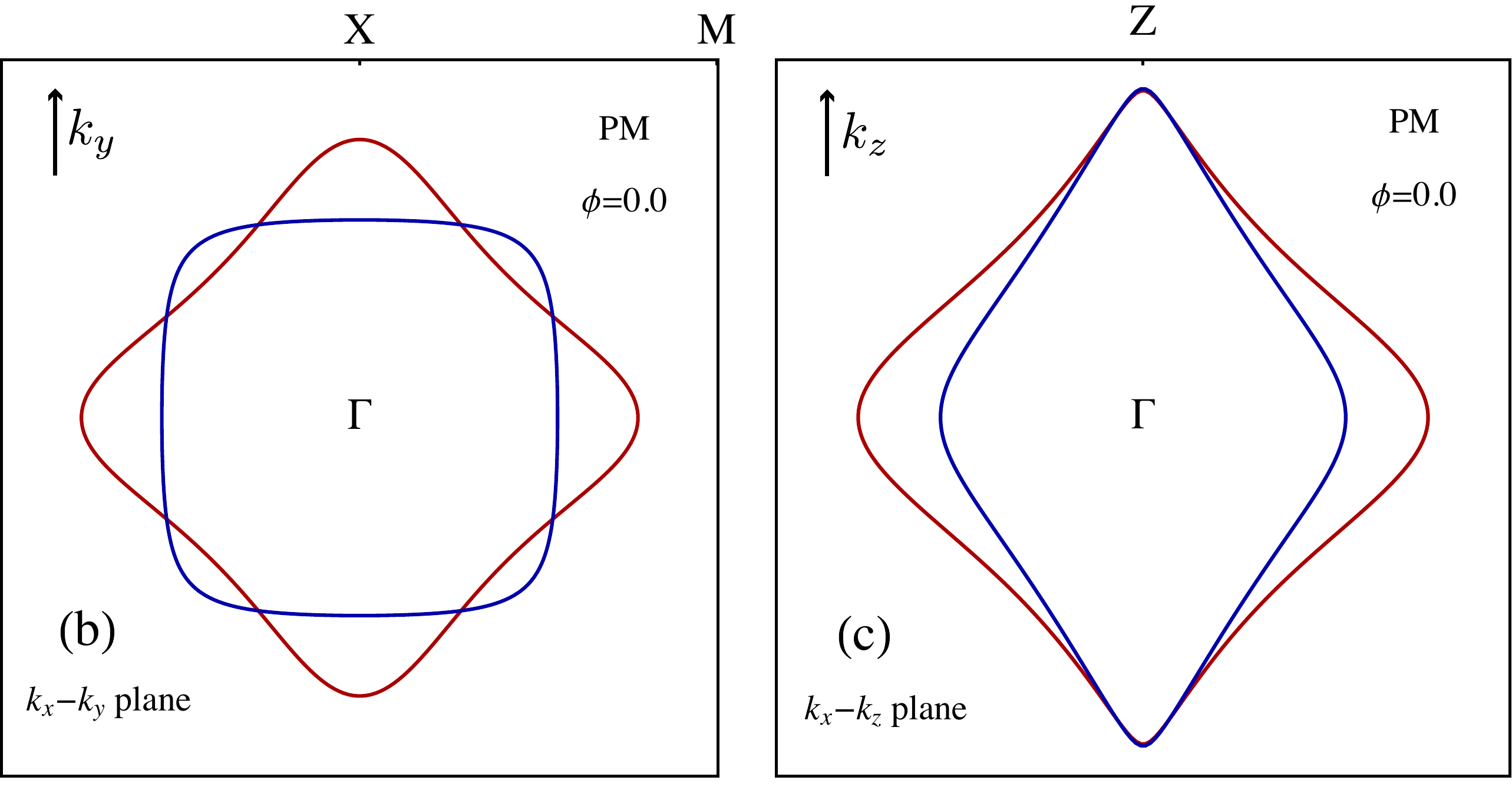}
\caption{
(Color online) 
(a) Fermi surface sheets of \URU~ in the extended simple tetragonal (st) BZ with electron sheet at $\Gamma (0,0,0)$ and hole sheet around $Z (0,0,2\pi/c)$ ($c\equiv 1$). The two sheets are nested by $\bQ = (0,0,1)$. FS cuts with $k_z=0$ (b) and $k_y=0$ (c)  in the reduced st BZ (Z-point folded onto $\Gamma$). Momentum range in (b) and (c) is given by $-\pi\leq k_{i}\leq \pi$.
\vspace{0.5cm}}
\label{fig:Fig1}
\end{figure*}

In this work we are interested in the very low energy ($\simeq 1$ meV) quasiparticle spectroscopy of \URU~therefore it is reasonable to start with an effective low energy model of the heavy electron bands. It should be simple enough to enable  analytical representation of the dispersion and high resolution computation of the QPI spectrum. But it must also have enough complexity to allow for modeling of realistic Fermi surface features, in particular the electron hole nesting property at wave vector $\bQ=(2\pi/c)\hbz=(0,0,1)$ in r.l.u. (reduced lattice units) because the latter leads to the staggered hidden order parameter. The kinetic energy may be constructed from hopping terms using the jj-coupled  single electron 5f states (j=total angular momentum) with incorporated spin-orbit (s.o.) coupling and orbital symmetries that are adapted to the local tetragonal crystalline electric field (CEF) potential at U sites. Such a procedure has been used successfully before in Refs.~(\onlinecite{takimoto:08,thalmeier:11,ikeda:12,thalmeier:13}). Since the s.o. splitting of U is very large the $j=7/2$ orbitals
are neglected, taking only the $j=5/2$ states which are CEF split into three Kramers doublets $\Gamma_7^{(\alpha)}$ $(\alpha=1,2)$ and $\Gamma_6$. For the formation of high order multipoles discussed here a further restriction to the two $\Gamma_7^{(\alpha)}$ doublets is possible. This is also suggested by ab-initio calculations of electron and hole pockets close to the Fermi surface \cite{oppeneer:10,ikeda:12}. These basis states are then created by $f^\dg_{\alpha\sigma}$ where $\sigma$ is the pseudo spin $(\si =\pm)$ of the doublets. They are related to the free ion states with total angular momentum component $j_z= M \; (|M|\leq 5/2)$ via the transformation
\bea
\left(
\begin{array}{c}
f_{1\pm}\\
f_{2\pm}
\end{array}
\right)
=
\left(
 \begin{array}{cc}
\cos\theta& \sin\theta \\
 -\sin\theta& \cos\theta
\end{array}
\right)
\left(
\begin{array}{c}
f_{\pm\frac{5}{2}}\\
f_{\mp\frac{3}{2}}
\end{array}
\right),
\label{eq:CEFtrans}
\eea
where the mixing angle $\theta$ is determined by the tetragonal CEF parameters. A minimal kinetic energy model involving CEF splitting and effective hopping up to second nearest neighbors (see appendix A) was introduced by Rau and Kee (Ref.~\onlinecite{rau:12}) and is given by
\be
\bl
H_0=&\sum_{\bk\si}\bigl(A_{1\bk}f^\dg_{1\si\bk}f_{1\si\bk}+A_{2\bk}f^\dg_{2\si\bk}f_{2\si\bk}\bigr)
+
\\&
\sum_\bk\bigl[D_{\bk}\bigl(f^\dg_{1+\bk}f_{2-\bk}-f^\dg_{2+\bk}f_{1-\bk}\bigr)
+H.c.
\bigr],
\label{eq:bands}
\el
\ee
%
%\end{widetext}
%
where the kinetic energy functions $A_{\al\bk}$ and $D_\bk$ are defined in Appendix \ref{sec:app1}. The above Hamiltonian directly parametrizes the heavy 5f quasiparticle states that form the electron and hole pockets. Therefore the hybridization with light electrons does not appear explicitly any more. In QPI spectroscopy this constrains us to the low energy region below the hybridization gap. The hybridization effects on local DOS were studied in Ref. ~\onlinecite{yuan:12}.
The model band structure and Fermi surface with electron pocket around $\Gamma$ point and hole pocket around Z (0,0,0.5) are shown in Fig.~\ref{fig:Fig1}. 

%
%%%%%%%%%%%%%%%%%%%%%%%%%%%%%%%%%%%%%%%%%%%%%%%%%%%%%%%%%%%%%%%%%%%%%
%%%%%%%%%%%%%%%%%%%%%%%%%%%%%%%%%%%%%%%%%%%%%%%%%%%%%%%%%%%%%%%%%%%%%
%

\section{The hidden multipolar order parameters}
\label{sec:multipolar}

The first step in the identification of spontaneous order, whether hidden or not, is the determination of  broken symmetries. 
In \URU~ these are \cite{thalmeier:13,shibauchi:14} {\it i}) translational symmetry breaking  due to the  antiferro-HO wave vector  \bQ~ (from band folding along $k_z$ observed in ARPES \cite{yoshida:13,meng:13}), {\it ii}) broken $C_4$ rotational symmetry (from torque oscillations \cite{okazaki:11}, cyclotron resonance splitting \cite{tonegawa:12} and high-resolution x-ray diffraction \cite{tonegawa:14}) and {\it iii}) time reversal symmetry breaking (from NMR \cite{takagi:12} and $\mu SR$ \cite{kawasaki:14}  experiments). It was concluded in Refs.~(\onlinecite{ikeda:12,thalmeier:13,shibauchi:14}) that the rank-5 dotriacontapole is the most plausible candidate. However, frequently the rank-4 hexadecapole was also proposed \cite{haule:09,kusunose:11} as a candidate, although it breaks only translational symmetry and reflectional in-plane symmetry but not time reversal. We will discuss both possibilities in this work.

First we give a prescription how to construct the multipolar order parameters and their molecular fields from the f- electron basis operators. It was demonstrated \cite{takimoto:08,thalmeier:13} that all physical f-electron multipoles up to highest rank 5 can be expressed in terms of the charge operator
\be
\rho_{\al\al'}=\frac{1}{2}\sum_\si f^\dg_{\al\si}f_{\al'\si},
\label{eq:pseudocharge}
\ee
and the pseudospin operator given by
\be
S^i_{\al\al'}=\frac{1}{2}\sum_{\si\si'}f^\dg_{\al\si}\si^i_{\si\si'}f_{\al'\si'},
\label{eq:pseudospin}
\ee
where generally $\alpha=1,2,3$ denote the $\Gamma^{(1),(2)}_7$ and $\Gamma_6$ orbitals, respectively and $\si^i$ denotes a Pauli matrix with Cartesian index $i=x,y,z$. Since we restrict to the former only multipoles with $\alpha,\alpha'=1,2$ can be constructed. The two candidates discussed here belong to this class. The explicit forms of the one- and two dimensional multipole representations at site $i$ are given by:
%%\begin{widetext}
%
\be
\bl
\mbox{hexadecapole, rank 4:}\;\;\;& \hat{\phi}_z^{A_{2+}}(i)=\frac{\oi}{\sqrt{2}}(S_{12}^z-S_{21}^z)_i
\el
\ee
and
\be
\bl
\mbox{dotriacontapole, rank 5:}\;
& 
\left\{
 \begin{array}{c}
\hat{\phi}_x^{E_-}(i)=
\frac{1}{\sqrt{2}}(S_{12}^x+S_{21}^x)_i
\\
%\;\; 
\hat{\phi}_y^{E_-}(i)=\frac{1}{\sqrt{2}}(S_{12}^y+S_{21}^y)_i
 \end{array}
 \right.
 .
\label{eq:OP}
\el
\ee
%%\end{widetext}
%
To clarify the meaning of the multipoles it is instructive to transform back to the free ion states using Eqs.~(\ref{eq:pseudospin},\ref{eq:CEFtrans}). For simplicity we choose the non-degenerate $A_{2+}$ state. For the Fourier transform 
$$ \hat{\phi}_z^{A_{2+}}(\bq)
=\frac{1}{N}\sum_i\exp(
\oi\bq \cdot \bR_i)\hat{\phi}_z^{A_{2+}}(i),$$
we obtain
\be
\bl
\hat{\phi}_z^{A_{2+}}(\bQ)=&\hat{\phi}_z^{A_{2+}}(-\bQ)^\dg
%^{}
\\
=&
-\oi
\kappa\frac{1}{N}\sum_\bk\sum_{|M|=\frac{3}{2},\frac{5}{2}}
\sigma_Mf^\dg_{M-4\si_M;\bk}f^{}_{M;\bk+\bQ}
\\
=&
\oi
\kappa\frac{1}{N}\sum_\bk
\bigl(f^\dg_{\fu\bk}f^{}_{-\fl\bk+\bQ}+f^\dg_{\fl\bk}f^{}_{-\fu\bk+\bQ}
\\
&
-f^\dg_{-\fu\bk}f^{}_{\fl\bk+\bQ}-f^\dg_{-\fl\bk}f^{}_{\fu\bk+\bQ}\bigr),
\el
\ee
where $\sigma_M=sign(M)=\pm 1$ and $\kappa=1/(2\sqrt{2})=0.35$. This explicit representation shows that the $A_{2+}$ hexadecapole (rank 4) is formed by condensation of electron-hole pairs with momenta (\bk, \bk+\bQ) in basis states which differ in angular momentum component by  $\Delta M=\pm 4$, leading to the high multipole property. The common antiferromagnetic order (rank 1) would be formed by pairs that differ by $\Delta M=\pm 1$.  Similar one may show that in the $E_-$ dotriakontapole (rank 5) is due to condensation of pairs with a maximum angular momentum difference $\Delta M=\pm 5$.
%
%%%%%%%%%%%%%%%%%%%%%%%%%%%%%%%%%%%%%%%%%%%%%%%%%%%%%%%%%%%%%%%%%%%%%
\begin{SCfigure*}
\includegraphics[width=1.4\linewidth,clip]{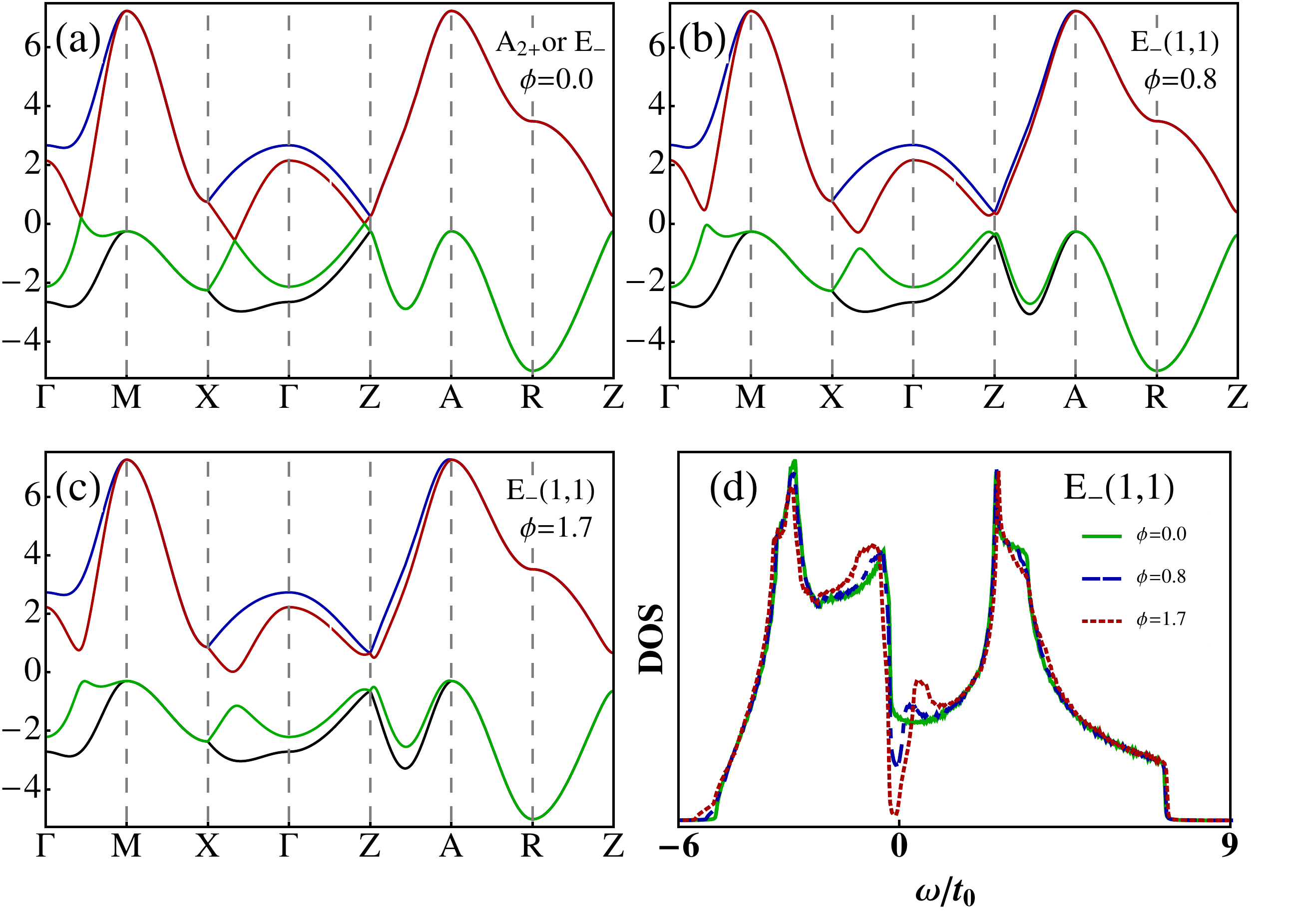}
\caption{
(Color online) 
Dispersion along st BZ path $\Gamma(0,0,0)$, $\mbox{M}(\frac{1}{2},\frac{1}{2},0)$;  $\mbox{X}(0,\frac{1}{2},0)$;  $\mbox{Z}(0,0,\frac{1}{2})$;  $\mbox{A}(\frac{1}{2},\frac{1}{2},\frac{1}{2})$;  $\mbox{R}(0,\frac{1}{2},\frac{1}{2})$.
(a) Effective f-bands in the para phase. (b, c) Reconstructed quasiparticle bands in the $E_-(1,1)$ HO phase for $\phi=0.8$ and $1.7$. The gapping of quasiparticle dispersion at \bk-points connected by the nesting vector $\bQ=(0,0,1)$ (r.l.u.)  can be clearly seen. (d) DOS with evolution of HO gap for small $\omega$ (also Fig.~\ref{fig:Fig4}c). Here and in subsequent figures we define $\phi = |\phi_z^\bQ|$ or $|\bphi^\bQ|$ in units of  $t_0=6.66$ meV.
\vspace{2cm}
}
\label{fig:Fig2}
\end{SCfigure*}
%%%%%%%%%%%%%%%%%%%%%%%%%%%%%%%%%%%%%%%%%%%%%%%%%%%%%%%%%%%%%%%%%%%%%
%
The effective interaction between f-quasiparticles leads to the instability in these multipole channels \cite{ikeda:12}. 
The ordered phase is then described by an additional molecular field term in the Hamiltonian controlled by the multipole
expectation values $\phi^\Gamma_n(\bQ)=\bra \hat{\phi}^\Gamma_n(\bQ)\ket$ where $\Gamma$ denotes the representation and $n$ is its degeneracy index. Using Eq.~(\ref{eq:pseudospin}) these terms may be written as
\be
\bl
& A_{2+}\!: H_\phi\!=\!
-\oi
\kappa\phi_z^\bQ\sum_\bk(f^\dg_{1\bk}\sigma_zf^{}_{2\bk+\bQ}-\! f^\dg_{2\bk}\sigma_zf^{}_{1\bk+\bQ}) \! +\!H.c. \\
 %\non\\
& E_{-}: H_\phi \!=-\kappa\bphi^\bQ\cdot\sum_\bk(f^\dg_{1\bk}\bsig f^{}_{2\bk+\bQ}+f^\dg_{2\bk}\bsig f^{}_{1\bk+\bQ}) +H.c.
\label{eq:HOP}
\el
\ee
Here we introduced $f^\dg_{\al\bk}=( f^\dg_{\al +\bk},f^\dg_{\al -\bk})$ with $\al=1,2$ and $\sigma_z=\pm$ denoting band index and Kramers pseudo-spin, respectively.  Furthermore $\bphi^\bQ=(\phi^\bQ_x,\phi^\bQ_y)$ is the HO vector which expresses the twofold degeneracy of the $E_{-}$ representation. Therefore right at $T_{HO}$ the HO phase has continuous $U(1)$ symmetry which is lifted by higher order terms in the free energy below $T_{HO}$. Commonly a phase with equal components $\phi^\bQ_x=\phi^\bQ_y$ called $E_-(1,1)$ phase or with only one nonzero $\phi^\bQ_x$ or $\phi^\bQ_y$ component called  $E_-(1,0)$ or  $E_-(0,1)$, respectively, is stabilized. In both cases different domains are possible. For a discussion of the domain issue we refer to Refs.(\onlinecite{thalmeier:11,thalmeier:13}).
A nonzero third component $\phi^\bQ_z$ of the HO vector $\bphi$
(replacing $x,y\rightarrow z$ in the last of Eq.~(\ref{eq:OP})) would correspond to a different $A_{2-}$ representation 
\cite{ikeda:12,rau:12} that will not be considered here. We introduce the $2\otimes4$ spinor basis $\Psi^\dg_\bk=(\psi^\dg_{a\bk},\psi^\dg_{b\bk})$ with effective Kramers degenerate a,b components defined below for each HO symmetry separately. The total mean field Hamiltonian, including the HO molecular fields may be written as
\be
H=H_0+H_\phi=\sum_\bk\Psi^\dg_\bk h_\bk \Psi_\bk;  \;\; h_{\bk}= h_{a\bk}\otimes h_{b\bk}.
\label{eq:HAMfac}
\ee
In the following we will diagonalize this Hamiltionian consisting of two $4\times 4$ blocks (a,b) explicitly in analytical form to calculated the reconstructed quasiparticle bands and the necessary Green's functions for QPI.

\section{Quasiparticle excitations in the HO phase}
\label{sec:HOquasiparticle}

The HO molecular fields result in a splitting of Fermi surface states connected by a nesting vector. This will reconstruct the Fermi surface and equal energy surfaces close to the hot spots of the nesting. The details of the FS reconstruction should leave its imprint in the QPI spectrum. Since the reconstruction depends on the symmetries of HO, the QPI characteristics may allow to discriminate between them, in the same way as it does for different gap symmetries in an unconventional superconductor.
For clarity this analysis  will be done separately for both HO candidates.

\subsection{dotriacontapole $E_-$ phase}
\label{subsec:dotria}

%
%%%%%%%%%%%%%%%%%%%%%%%%%%%%%%%%%%%%%%%%%%%%%%%%%%%%%%%%%%%%%%%%%%%%%
\begin{figure*}
\includegraphics[width=50mm,clip]{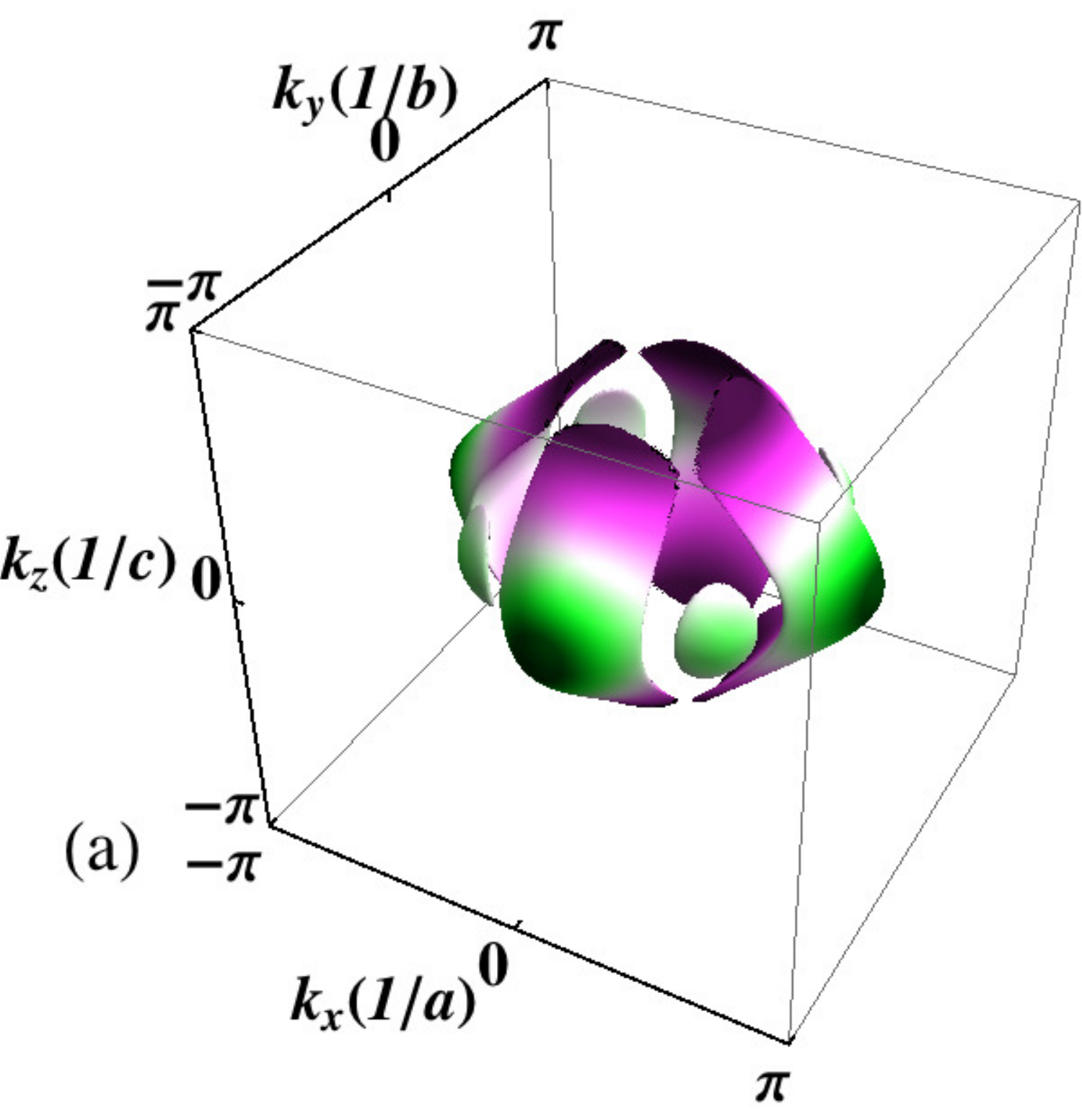}\hspace{0.3cm}
\includegraphics[width=95mm,clip]{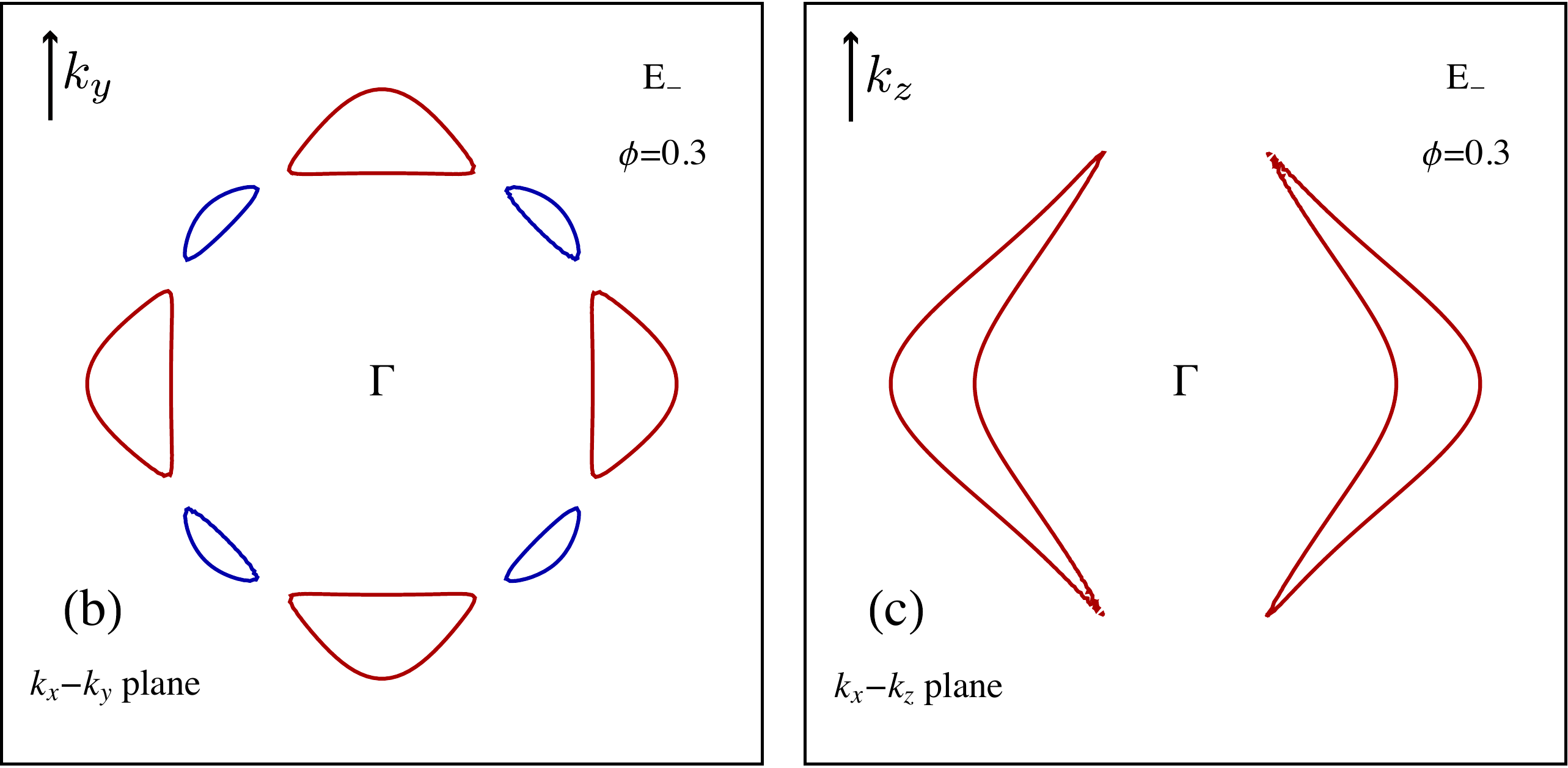}
\caption{
(Color online) 
(a) Fermi surface sheets of \URU~in the $E_-(1,1)$ HO phase in the reduced BZ (folded by \bQ). The FS is reconstructed in \bk -space regions connected by the nesting vector $\bQ = (0,0,1)$ and breaks up into four larger and four smaller sheets (partly hidden). FS cuts in  (b) the $k_x-k_y$ plane and (c) the $k_x-k_z$ plane.
Momentum range in (b) and (c) is given by $-\pi\leq k_{i}\leq \pi$.
%\vspace{0.5cm}
}
\label{fig:Fig3}
\end{figure*}
%%%%%%%%%%%%%%%%%%%%%%%%%%%%%%%%%%%%%%%%%%%%%%%%%%%%%%%%%%%%%%%%%%%%%
%

In this representation it is most convenient to express the the Hamiltonian in the spinor basis 
\be
\bl
\psi^\dg_{a\bk}=(f^\dg_{1+\bk},f^\dg_{2-\bk},f^\dg_{1+\bk+\bQ},f^\dg_{2-\bk+\bQ}),
%;\;\;
%\nonumber
\\
\psi^\dg_{b\bk}=(f^\dg_{1-\bk},f^\dg_{2+\bk},f^\dg_{1-\bk+\bQ},f^\dg_{2+\bk+\bQ}),
\label{eq:NambuE}
\el
\ee
where it factorizes into two Kramers degenerate $(a,b)$ $4\otimes 4$ blocks.
The generally two component order parameter is given by $\bphi=(\phi_x,\phi_y)$ (with  ordering wave vector \bQ~ now suppressed).  The resulting quasiparticle energies are the eigenvalues of $\tilh_\bk=(h_{\bk}-\omega I)=\tilh_{a\bk}\otimes\tilh_{b\bk}$.
The $4\otimes 4$ Hamiltonian blocks in spinor basis are given by
\be
\bl
&
\tilh_{a\bk}=
\left(
 \begin{array}{cc}
\ha_\bk& \hla_a\\
 \hla_a& \ha_{\bk+\bQ}
\end{array}
\right);\;\;\;
\ha_{\bk}=
\left(
 \begin{array}{cc}
\tA_{1\bk}& D_\bk\\
D_\bk^*& \tA_{2\bk}
\end{array}
\right);
%\;\;\;\nonumber
\\\\&
\hla_a=
\left(
 \begin{array}{cc}
0&-\ka(\phi_x-
\oi
\phi_y)\\
-\ka(\phi_x+
\oi
\phi_y) & 0
\end{array}
\right),
\label{eq:HblocksE}
\el
\ee
with $\tA_{\al\bk}=A_{\al\bk}-\omega$  where $A_{\al\bk}$ and $D_\bk$ are defined in Appendix A. The secular equation   $|\tilh_{a\bk }|=0$ for the eigenvalues  is obtained as
%%
%\begin{widetext}
%
\be
\bl
&
|D_\bk|^2(\tA_{1\bk}\tA_{2\bk}+\tA_{1\bk+\bQ}\tA_{2\bk+\bQ})
\\
&
-\bigl[(\tA_{1\bk}\tA_{2\bk+\bQ}-\ka^2|\bphi|^2)(\tA_{2\bk}\tA_{1\bk+\bQ}-\ka^2|\bphi|^2) + |D_\bk|^4\bigr]
\\
&
-2\ka^2|\bphi|^2\tiD_\bk
=0,
\label{eq:secuE}
\el
\ee
%%
%\end{widetext}
%
where we defined the real function 
$$\tiD_\bk=\frac{1}{2}(c^2D^2_\bk+c^{*2}D^{*2}_\bk),$$
 and $c=(\hphi_x+\oi\hphi_y)$ with $|c|^2=1$.
Solving this fourth order equation leads to the closed solution for the four HO quasiparticle bands $(i=1-4)$ valid for general \bk~($\pm$ chosen idependently):%\\
\be
\bl
\vare_{i\bk}=&\vare^\pm_{1,2}(\bk)=A_\bk^\perp\pm(\omega_0^2\pm\tilde{\omega}_0^2)^\frac{1}{2},
\\
\omega^2_0=&A_\bk^{z2}+\Delta_\bk^{\perp 2} +|D_\bk|^2+\kappa^2|\bphi |^2,
\\
\tilde{\omega}^2_0=&2\bigl[A_\bk^{z2}(\Delta_\bk^{\perp 2}+|D_\bk|^2)+\kappa^2|\bphi |^2\zeta_\bk\bigr]^\frac{1}{2},
\label{eq:dispE0}
\el
\ee
or, explicitly $(i=\pm, 1,2)\equiv (i=1-4)$
\be
\bl
\vare_{i\bk}=\vare^\pm_{1,2}(\bk)=A_\bk^\perp\pm
\{ (A_\bk^{z2}+\Delta_\bk^{\perp 2} +|D_\bk|^2+\kappa^2|\bphi |^2)
\\
\pm 2\bigl[A_\bk^{z2}(\Delta_\bk^{\perp 2}+|D_\bk|^2)+\kappa^2|\bphi |^2\zeta_\bk\bigr]^\frac{1}{2}\}^\frac{1}{2},
\label{eq:dispE}
\el
\ee
\\
where on the l.h.s. $\pm$ corresponds to the second and $1,2$ to the first $\pm$ on the r.h.s., respectively. 
Here we defined $|\bphi|=(\phi_x^2+\phi_y^2)^\frac{1}{2}$ as the order parameter amplitude  with $\hat{\bphi}=\bphi/|\bphi|= (\hphi_x,\hphi_y)$ and $A_\bk^{z,\perp}$, $\Delta_\bk^\perp$ and $D_\bk$ are given in Appendix \ref{sec:app1}. Furthermore, with $D_\bk=D'_\bk+iD''_\bk$ we introduce the azimuthal function $\zeta_\bk$ which leads to the breaking of fourfold $C_4$ symmetry in the HO phase. Its general form is given by
\be
\bl
\zeta_\bk
=&\frac{1}{2}(|D_\bk|^2-\tiD_\bk)
=
%\\&
\frac{1}{2}\bigl[{D'_\bk}^2 +{D''_\bk}^2+(D'_\bk\hphi_y+D''_\bk\hphi_x)^2
\\
&\hspace{3cm}
- (D'_\bk\hphi_x-D''_\bk\hphi_y)^2\bigr].
\label{eq:zeta}
\el
\ee
In the context of a Landau expansion of the free energy below T$_{HO}$,
 it  is concluded that only phases $\hat{\bphi}=(\hphi_x,\hphi_y)$ are stable where both components have the same modulus  $(\hphi_x,\pm\hphi_y)=\frac{1}{\sqrt{2}}(1,\pm 1)$ denoted by $E_-(1,1)$ and $E_-(1,\bar{1})$ or one of the component vanishes  $(\hphi_x,\hphi_y)=(1,0), (0,1)$, denoted by $E_-(1,0)$ and  $E_-(0,1)$. In each case the two possibilities are two different domains of the same phase. The single component phase has been ruled out by torque experiments \cite{okazaki:11,thalmeier:11} therefore we will only consider the two component phase in the following. We get
\be
\bl
E_-(1,1):          \zeta_\bk  = &\frac{1}{2}(D'_\bk+D''_\bk)^2  
\\
= & 32t_{12}^2\cos^2\fa k_x\sin^2\fa k_y\sin^2\fc k_z ,
\\
E_-(1,\bar{1}): \zeta_\bk  = & \frac{1}{2}(D'_\bk-D''_\bk)^2 
\\
= & 32t_{12}^2\sin^2\fa k_x \cos^2\fa k_y\sin^2\fc k_z .
\label{eq:zeta2}
\el
\ee
This (positive) function breaks fourfold $C_4$ rotational symmetry under a coordinate rotation  $\bk\rightarrow\bk'$ by $\pi/2$ with $k'_x = k_y$, $k_y'=-k_x$. Therefore the in-plane symmetry is reduced to twofold rotations $C_2$ generally since $\zeta_{-\bk}=\zeta_\bk$.
The different domains correspond to a $\pi/2$ rotation of \bk~ and therefore to a  relative $\pi/2$- shift of the fourfold symmetry breaking effects  \cite{thalmeier:11}. In the following we will mostly discuss the $E_-(1,1)$  domain of the two component phase. We note that the $C_4$ symmetry breaking, due to a nonvanishing $\zeta_\bk$, is directly tied to the inter-orbital hopping $t_{12}$,
 because the order parameter is an inter-orbital electron hole condensate. Therefore it is absent for $k_z$=0.\\

%
%%%%%%%%%%%%%%%%%%%%%%%%%%%%%%%%%%%%%%%%%%%%%%%%%%%%%%%%%%%%%%%%%%%%%
\begin{figure*}
%\vspace{0.2cm}
{\includegraphics[width=.9\linewidth,clip]{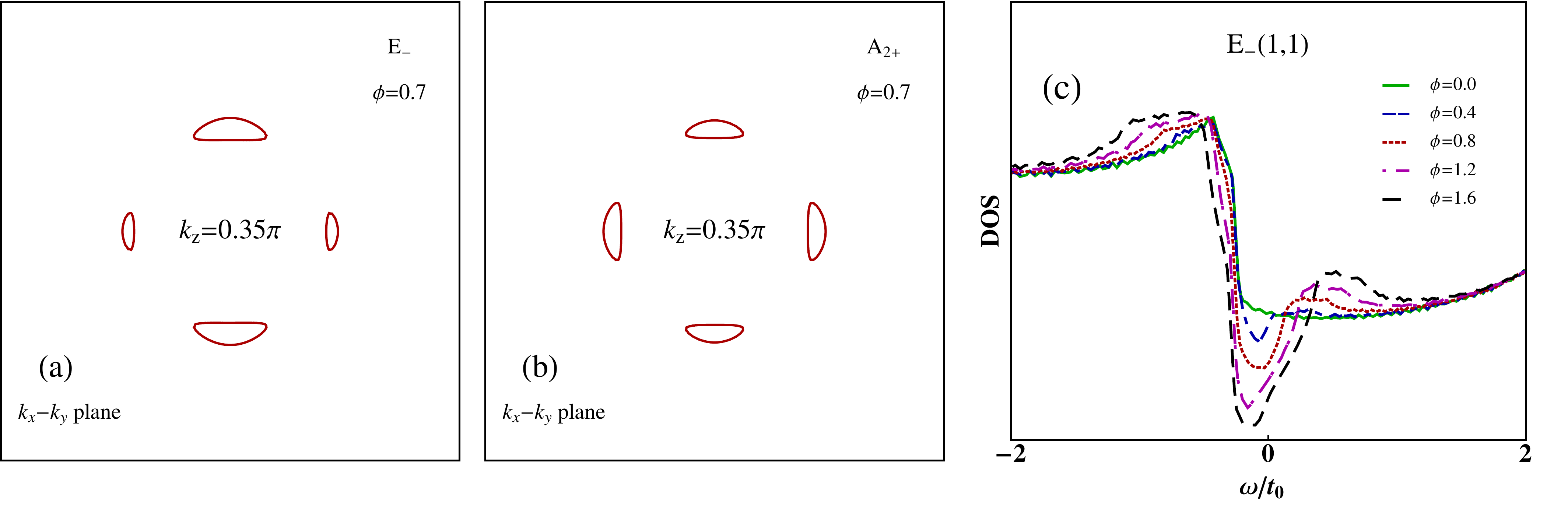}}
%\vspace{0.5cm}
\caption{
(Color online) 
Comparison of reconstructed Fermi surface sheets of \URU~ for different HO symmetry in the $k_x-k_y$ plane with $|k_z|=0.35\pi$. (a) In the $E_-(1,1)$ phase the breaking of fourfold rotational $C_4$ symmetry for $k_z\neq 0$ is obvious. Changing to  $E_-(1,\bar{1})$ domain corresponds to $\pi/2$ rotation. (b) For $A_{2+}$ HO fourfold symmetry is preserved. (c) Zoomed DOS in the HO gap region for various $\phi$. Charge carrier DOS is reduced to small values for large $\phi$.
Momentum range in (a) and (b) is given by $-\pi\leq k_{i}\leq \pi$.
}
\label{fig:Fig4}
\end{figure*}
%%%%%%%%%%%%%%%%%%%%%%%%%%%%%%%%%%%%%%%%%%%%%%%%%%%%%%%%%%%%%%%%%%%%%

The second $4\times 4$ block in Eq.~(\ref{eq:HAMfac}), $\tilh_{b\bk}$, is obtained from $\tilh_{a\bk}$ by replacing $\ha_\bk\rightarrow\hb_\bk$ as obtained from $D_\bk\rightarrow-D_\bk^*$ and in addition replacing $\hla_a\rightarrow\hla_a^\dg=\hla_b$.  Since  $|\tilh_{b\bk}| \equiv  |\tilh_{a\bk}|$ the resulting quasiparticle dispersion from  $|\tilh_{b\bk}|=0$ is identical to $\omega^\pm_{1,2}(\bk)$ in Eq.~(\ref{eq:dispE}). Therefore, altogether, each of these four branches (due to two orbitals and the unit cell doubling by ordering vector \bQ) is in addition twofold degenerate. This degeneracy is due to the invariance under combined time reversal and translation by \bQ, therefore the $a,b$ equivalence  is an effective Kramers degeneracy not lifted by the antiferro-type order, although time reversal symmetry itself is broken.\\

\subsection{hexadecapole $A_{2+}$ phase}
\label{subsec:hexa}

The order parameter $\phi_z$ of this phase (Eq.~(\ref{eq:HOP})) is non-degenerate. Similar to previous case the 
full Hamiltonian may be given in block form as in Eq.~(\ref{eq:HblocksE}), but with a different spinor basis and  molecular field part:
We now use the reordered basis (the Kramers index in the last pair in a,b is interchanged):
\bea
\psi^\dg_{a\bk}=(f^\dg_{1+\bk},f^\dg_{2-\bk},f^\dg_{1-\bk+\bQ},f^\dg_{2+\bk+\bQ}),
%;\;\;
\nonumber\\
\psi^\dg_{b\bk}=(f^\dg_{1-\bk},f^\dg_{2+\bk},f^\dg_{1+\bk+\bQ},f^\dg_{2-\bk+\bQ}),
\label{eq:NambuA2}
\eea
then we obtain 
\be
\bl
&
\tilh_{a\bk}=
\left(
 \begin{array}{cc}
\ha_\bk& \hla_a\\
 \hla^\dg_a& \ha^*_{\bk+\bQ}
\end{array}
\right);\;\;\;
\ha_{\bk}=
\left(
 \begin{array}{cc}
\tA_{1\bk}& D_\bk\\
D_\bk^*& \tA_{2\bk}
\end{array}
\right);
%;\;\;\;
\\
\\&
\hla_a=
\left(
 \begin{array}{cc}
0&-i\ka\phi_z\\
-i\ka\phi_z & 0
\end{array}
\right).
\label{eq:HblocksA}
\el
\ee
The second block $\tilh_{b\bk}$ is obtained from  $\tilh_{a\bk}$ by replacing $\ha_\bk\rightarrow \hb_\bk$ through $D_\bk\rightarrow -D^*_\bk$ and in addition $\hla_a\rightarrow -\hla_a=\hla_b$. The secular equation  $|\tilh_{a\bk }|=0$ is similar to Eq.~(\ref{eq:secuE}):
%
%\begin{widetext}
%
\be
\bl
&
|D_\bk|^2(\tA_{1\bk}\tA_{2\bk}+\tA_{1\bk+\bQ}\tA_{2\bk+\bQ})-
\\&
\bigl[(\tA_{1\bk}\tA_{2\bk+\bQ}-\ka^2|\phi_z|^2)(\tA_{2\bk}\tA_{1\bk+\bQ}-\ka^2|\phi_z|^2) + |D_\bk|^4\bigr]
\\&
-2\ka^2|\phi_z|^2|D_\bk|^2=0.
\label{eq:secuA2}
\el
\ee
%\end{widetext}
Again the four quasiparticle bands $(i=1-4)$ in the $A_{2+}$ HO phase may be obtained in closed form as
\be
\bl
&
\vare_{i\bk}=\vare^\pm_{1,2}(\bk)=A_\bk^\perp\pm(\omega_0^2\pm\tilde{\omega}_0^2)^\frac{1}{2},
\\&
\omega^2_0=A_\bk^{z2}+\Delta_\bk^{\perp 2} +|D_\bk|^2+\kappa^2|\phi_z |^2,
\\&
\tilde{\omega}^2_0=2|A^z_\bk |\bigl[\Delta_\bk^{\perp 2}+|D_\bk|^2\bigr]^\frac{1}{2},
\label{eq:dispA20}
\el
\ee
or, explicitly $(i=\pm, 1,2)\equiv (i=1-4)$
\be
\bl
&\vare_{i\bk}=E^\pm_{1,2}(\bk)=A_\bk^\perp\pm
 \\
&
\sqrt{A_\bk^{z2}+\Delta_\bk^{\perp 2} +|D_\bk|^2+\kappa^2|\phi_z|^2
\pm
 2|A_\bk^{z}|
 \sqrt{\Delta_\bk^{\perp 2}+|D_\bk|^2}
}.
\label{eq:dispA2}
\el
\ee
%
%\\
%
The four branches from $|\tilh_{b\bk} |=0$ are identical to those of  $|\tilh_{a\bk} |=0$ leading to a Kramers degeneracy similar as before (For $A_{2+}$ time reversal symmetry is already preserved by itself). 
The result in  Eq.~(\ref{eq:dispA2}) is obtained from the $E_-$ case dispersion of Eq.~(\ref{eq:dispE}) formally by replacing $|\bphi|\rightarrow |\phi_z|$ and $\zeta_\bk\rightarrow 0$. Therefore, due to the non-degeneracy of $A_{2+}$ which implies $|\phi_z|^2$ transforming like $A_{2+}\otimes A_{2+}=A_{1+}$,  there is no term in the 
dispersion that breaks the fourfold rotational symmetry (and also not the in-plane reflection symmetry). This should give an important distinction in the QPI spectrum of the two phases.\\
%
%%%%%%%%%%%%%%%%%%%%%%%%%%%%%%%%%%%%%%%%%%%%%%%%%%%%%%%%%%%%%%%%%%%%%
\begin{SCfigure*}
%\vspace{0.2cm}
\includegraphics[width=1.4\linewidth,clip]{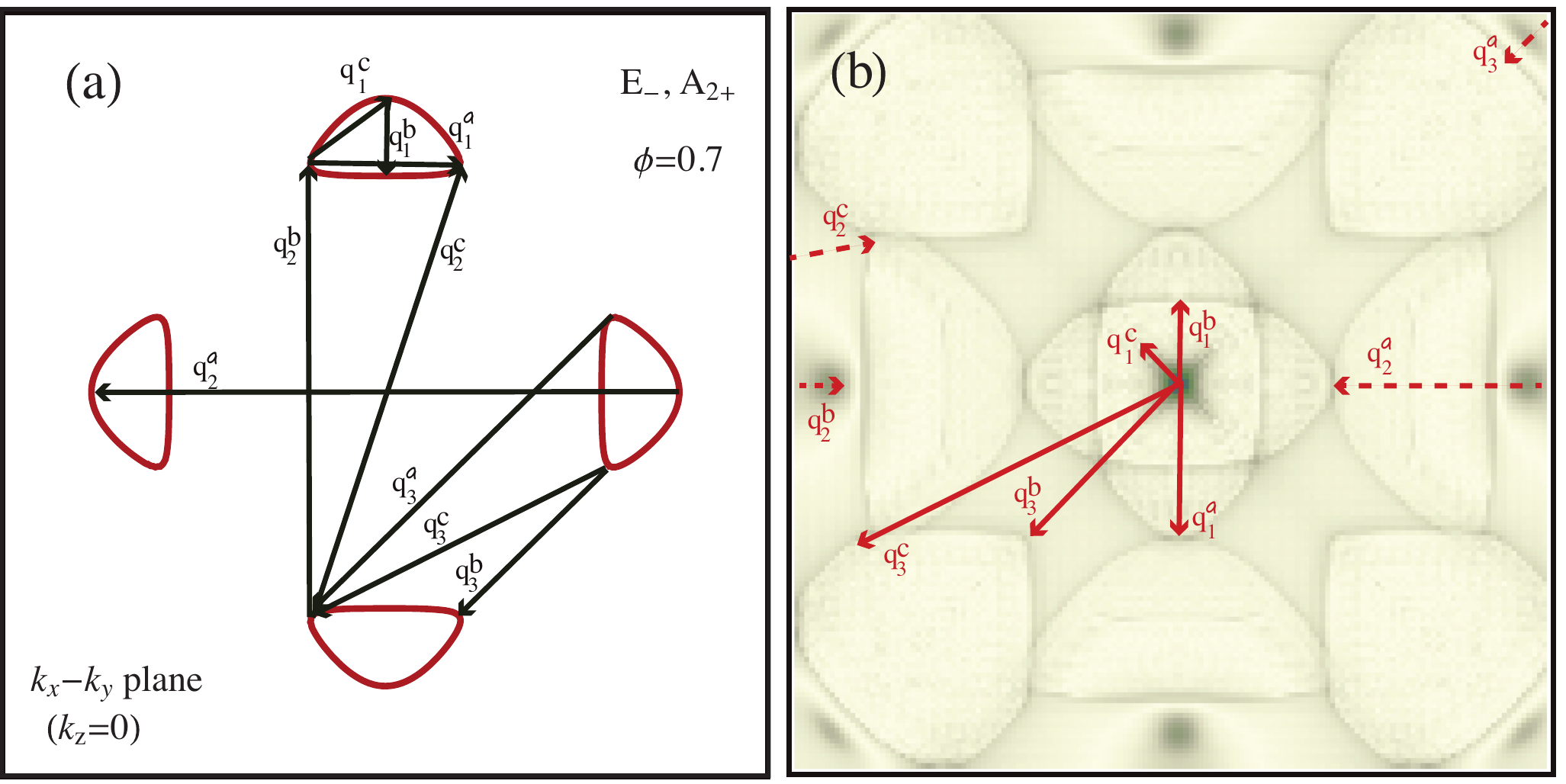}
%\vspace{-0.5cm}
\caption{
(Color online) 
(a) Reconstructed HO Fermi surface ($\omega=0$) for $\phi=0.7$ (in units of $t_0$) in the $k_x-k_y$ plane with $k_z=0$ where $E_-,A_{2+}$ are equivalent. The characteristic scattering wave vectors $\bq_{1-3}^{a-c}$ %,\bq_2^{a-c},\bq_3^{a-c}$ 
for the QPI spectrum are indicated. (b) Partial QPI (absolute value) spectrum $(\omega =0$) for $k_z=0$ slice of the Fermi surface. All $\bq_i^\alpha$ are present and the image of HO FS sheets that have doubled `$2k_F$' dimension is clearly visible.
Momentum range  is given by $-\pi\leq k_{i}\leq \pi$. Dashed arrows denote image folded back into the first BZ.
\vspace{0.4cm}
}
\label{fig:Fig5}
\end{SCfigure*}
%%%%%%%%%%%%%%%%%%%%%%%%%%%%%%%%%%%%%%%%%%%%%%%%%%%%%%%%%%%%%%%%%%%%%
%

\subsection{Dispersion in special cases for $E_-$ and $A_{2+}$}
\label{subsec:special}

We also discuss some special and limiting cases for the dispersion for greater clarity: In the $E_-$ phase we have
%%\begin{widetext}
%
\be
\bl
\bphi=0 \;\;\;\;\;
&
:\;
\vare^\pm_{1,2}(\bk)=\!
A_\bk^\perp\pm A^z_\bk \pm
\sqrt{\De^{\perp 2}_\bk+|D_\bk|^2},
\\
D_\bk
\!
=
\!
0\; 
(k_z
\!
=
\!0)
&
:\;
\vare^\pm_{1,2}(\bk)=\!\!
A_\bk^\perp\pm
%\bigl[
\!
\sqrt{
(A^z_\bk\pm\De^\perp_\bk)^2+
\!
\kappa^2|\bphi|^2},
%\bigr]^\frac{1}{2}
\\
D_\bk
\!
=
\!
0
;\; \bphi=0
\;
&
:\;
\vare^\pm_{1,2}(\bk)=\!
A_\bk^\perp
\pm A_\bk^z\pm\De_\bk^\perp.
\label{eq:special}
\el
\ee
%%\end{widetext}
%
Note that the $\pm$ signs are chosen in arbitrary combination to give four bands (which are in addition twofold Kramers degenerate).
The first equation describes quasiparticle bands in the para phase for general \bk, the second in the ordered phase for in-plane wave vector and the last one for both conditions satisfied. Taking into account $A^z_{\bk+\bQ}=-A^z_\bk$ the second equation is equivalent 
to the result in Ref.~\onlinecite{rau:12} in the Brillouin zone of the ordered phase. 
These special cases are described by identical dispersions  for the $A_{2+}$ phase when we replace $|\bphi|\rightarrow \phi_z$ in the second equation. The other equations refer to the para phase. Due to this identity the reconstructed HO FS cuts with $k_z=0$ are the same for both phases because then $\zeta_\bk=0$ for $E_-$. In order to see the difference between $E_-$ and $A_{2+}$ we have to consider cuts with $|k_z| > 0$ where the fourfold symmetry breaking through nonzero  $\zeta_\bk$ appears.

\section{Green's function, quasiparticle DOS and QPI spectrum in HO phase}
\label{sec:QPI-HO}

To calculate the QPI spectrum for both HO models we need the Green's function $G_\bk=G_{a\bk}\otimes G_{b\bk}$ with  $G_{\ga\bk}=(\om 1-h_{\ga\bk})^{-1}$ ($\ga=a,b$). The Green's function matrices will be diagonal in the basis  of the four quasiparticle eigenvectors $(i,j=1,4)$ of $h_\bk^{a,b}$ which form the columns of the unitary transformation $U_{\ga\bk}$ in
\bea
h'_{\ga\bk}&=&U_{\ga\bk}h_{\ga\bk}U^\dagger_{\ga\bk};\;\;\
\{h'_{\ga\bk}\}_{ij}=\vare_{i\bk}\delta_{ij},
\label{eq:hdiagonal}
\eea
where  $\vare_{i\bk}$ has Kramers degeneracy with respect to $\ga$.
The primed spinors $\psi'_{\ga\bk}=U_{\ga\bk}\psi_{\ga\bk}$ corresponding to the eigenvectors satisfy the canonical anti-commutation relations. In this primed basis we obtain the Green's functions
\be
G'_{\ga\bk}(\om)=(\om 1-h'_{\ga\bk})^{-1},
\ee
therefore we have
\be
 G'_{\ga\bk}(\om)_{ij}=\frac{\delta_{ij}}{(\om -\vare_{i\bk})},
\label{eq:Green}
\ee
where $\vare_{i\bk}$ are the exact solutions for the quasiparticle bands in Eq.~(\ref{eq:dispE}) or Eq.~(\ref{eq:dispA2}). The Green's function does not depend on $\ga$. Then we may obtain the quasiparticle DOS (per Kramers pseudo spin degree) as
\be
\bl
N(\omega)
&=-\frac{1}{\pi} Im \frac{1}{N}
\sum_{i\bk}\Bigl(\frac{1}{\om-\vare_{i\bk}+\oi\eta}\Bigr)_{\om\rightarrow \omega+\oi\eta}
%\non
\\
&=
-\frac{1}{\pi} Im \frac{1}{N}
\sum_\bk\frac{4\omega_\bk(\omega_\bk^2-\omega_0^2)}{(\omega_\bk^2-\omega_0^2)^2-(\tilde{\omega}_0^2)^2},
\label{eq:DOS}
\el
\ee
where the sum over band index $i$ has been carried out explicitly in the second expression. Here $\omega_\bk=\omega-A^\perp_\bk +\oi\eta$ with $\eta\rightarrow 0$. The evolution of the HO gap in $N(\omega)$ is shown in Fig.~\ref{fig:Fig2}d.

 %
%%%%%%%%%%%%%%%%%%%%%%%%%%%%%%%%%%%%%%%%%%%%%%%%%%%%%%%%%%%%%%%%%%%%%
\begin{SCfigure*}
%\vspace{0.2cm}
{\includegraphics[width=1.5\linewidth,clip]{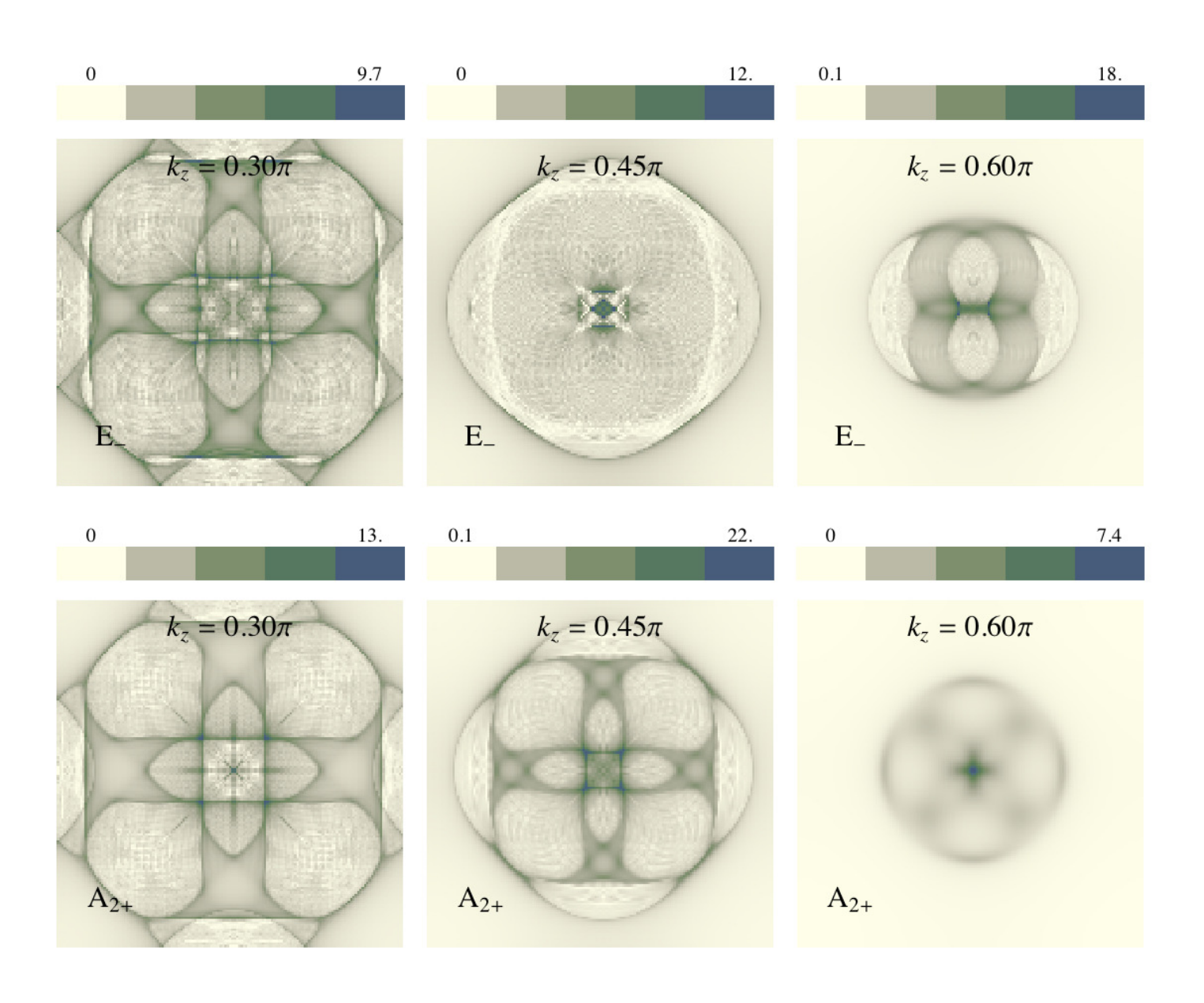}}
\caption{
(Color online) 
Partial QPI (absolute value) spectra ($\omega=0.124t_0$) for different $|k_z| > 0$ slices for $E_-$ (first row) and $A_{2+}$ (second row) HO. For larger $k_z$ distinct $C_4$ symmetry breaking due to the $\zeta_\bk$ function in the dispersion of Eq.~(\ref{eq:dispE}) appears for $E_-$ while $C_4$ is preserved for $A_{2+}$.
(Momentum range  is given by $-\pi\leq q_{x,y}\leq \pi$).
\vspace{4.75cm}
}
\label{fig:Fig6}
\end{SCfigure*}
%%%%%%%%%%%%%%%%%%%%%%%%%%%%%%%%%%%%%%%%%%%%%%%%%%%%%%%%%%%%%%%%%%%%%
%
The quasiparticle interference spectrum is the Fourier transform of the change in local density of states introduced by scattering from dilute impurities on the surface \cite{capriotti:03}. It may be calculated within t-matrix theory of the scattering. We assume a scattering potential strength $V_0\ll\phi$ that is small compared to the hidden order gap (Sec.~\ref{sec:numerical}). Then it is reasonable to express the impurity Hamiltionian directly within the basis of quasiparticles in the HO state. We take the simplest form of a non-magnetic momentum- and orbital- independent scattering excluding interband processes. In the primed (HO) quasiparticle basis it is written as
\bea
H_{imp}=\frac{V_0}{N}\sum_{\bk\bq}\bigl[{\psi'}^\dagger_{a\bk+\bq}{\psi'}^\dagger_{a\bk}+
{\psi'}^\dagger_{b\bk+\bq}{\psi'}^\dagger_{b\bk}\bigr].
\label{eq:scatt}
\eea
For  $\omega\ll\phi$ and sufficiently small $V_0$ as defined above the QPI spectrum may be treated  in Born approximation for the scattering  \cite{akbari:13,*akbari:13b} leading to a local DOS modification 
(per Kramers pseudo spin) given by:
\be
\delta N(\bq,\omega)
\!=\!
\frac{-V_0}{2\pi N} 
Im% \frac{1}{2N}
\!
\sum_{\ga\bk}
\!
 tr
\!
\bigl[G'_{\ga\bk}(\om) G'_{\ga\bk-\bq}(\om)\bigr]_{\om \!\rightarrow\omega +\oi\eta}
\label{eq:LDOS}
\ee
The trace may easily be evaluated and defining  $\delta N(\bq,\omega)=V_0\Lambda(\bq,\omega)$ we obtain the final QPI spectrum as:
\be
\Lambda(\bq,\om)=-\frac{1}{\pi} Im \frac{1}{N}\sum_{i\bk}
[\om-\vare_{i\bk}]^{-1}[\om-\vare_{i\bk-\bq}]^{-1}.
\label{eq:QPI_HO}
\ee
This expression for the QPI spectrum is valid for both $E_-$ and $A_{2+}$ type HO, using the quasiparticle dispersions 
$\vare_{i\bk}$ given in Eqs.~(\ref{eq:dispE},\ref{eq:dispA2}) for $E_-$ and $A_{2+}$ HO, respectively.

The \bq~ -vector in $\Lambda(\bq,\om)$ is a 2D surface vector. However, the integration over \bk~ has to be performed over the full 3D BZ of \URU~ since its FS has a 3D character and the $k_z$ component is not preserved in tunneling due to the surface. Nevertheless as shown in the example of the $d_{x^2-y^2}$-wave superconductor \CC\cite{akbari:11} it is instructive to consider the QPI spectrum for each $k_z$ slice of the FS separately for the presence of characteristic features of the equal quasiparticle energy surface in the HO phase. Then the summation over $k_z$ is performed  to see which of those features survive in the total QPI spectrum observed in experiment.

\section{QPI in the chiral superconducting phase embedded in hidden order}
\label{sec:QPI-SC}

At T$_c=1.45$ K , far below T$_{HO}=17.5$ K, \URU~ becomes superconducting. As a function of pressure this embedding in the HO phase is maintained up to the critical pressure of $p_c\simeq 0.7$ kbar where both superconductivity and hidden order vanishes \cite{amitsuka:07}. Therefore one may conjecture that HO is a necessary condition for SC to appear in this compound. Various experiments like field-angle dependent thermal conductivity \cite{kasahara:07,kasahara:09} and specific heat \cite{yano:08} measurements have been interpreted in terms of a chiral-d wave gap symmetry that has line and point nodes. Since the investigation of gap structures in heavy fermion-superconductors by QPI has recently been successfully demonstrated for \CC~ \cite{akbari:11,allan:13,zhou:13} it is worthwhile to perform a theoretical analysis of the predicted QPI pattern in the proposed chiral-d wave gap. Sofar, experimentally the QPI investigations \cite{schmidt:10} have been limited to the hidden order phase for $T_c < T \ll  T_{HO}$.

We start from the reasonable assumption $(T_c \ll T_{HO})$ that the SC order parameter is formed by pairing of reconstructed quasiparticles of the HO phase whose dispersion is given by Eqs.~(\ref{eq:dispE},\ref{eq:dispA2}). 
They are described  by the Nambu spinors ${\Psi'}^{i\dg}_{\bk}=({\psi'}^{i\dg}_{a\bk},{\psi'}^{i}_{b\bbk})$ with the definition $\bbk=-\bk$ and with $i=1-4$ denoting one of the four HO quasiparticle bands which are twofold $(a,b)$ Kramers degenerate.
Therefore indices $a\bk$ and $b\bbk$ refer to time reversed states with opposite quasiparticle Kramers pseudospin $(a,b)$ and momenta $(\bk,\bbk)$. The mean field Hamiltionian for singlet pairing of the effective pseudo-spin states of HO quasiparticles is then 
\be
\bl
H_{MF}=&\sum_{\bk i}
\left(
 \begin{array}{cc}
{\psi'}^{i\dg}_{a\bk} &{\psi'}^{i}_{b\bbk}
\end{array}
\right)
\left(
 \begin{array}{cc}
\vare_{i\bk}& \De_{i\bk}\\
 \De_{i\bk}^*& -\vare_{i\bk}
\end{array}
\right)
\left(
 \begin{array}{c}
{\psi'}^{i}_{a\bk} \\
{\psi'}^{i\dg}_{b\bbk}
\end{array}
\right)
\\
=&
\sum_{\bk i}{\Psi'}^{i\dg}_{\bk}\hh^i_\bk{\Psi'}^i_\bk,
\el
\ee
where $\De_{i\bk}$ is the singlet gap function discussed below.
The Green's function matrix is then given by $\hG_0^i(\bk,\om)=[\om-\hh^i_\bk]^{-1}$ which has the normal (diagonal) and anomalous (off-diagonal condensate) elements
\be
\bl
G^i_0(\bk,\om)
\!=\!
\frac{\om+\vare_{i\bk}}{(\om)^2-E^{2}_{i\bk}},
\\
F^i_0(\bk,\om)
\!=\!
\frac{\De_\bk}{(\om)^2-E^{2}_{i\bk}}.
\el
\ee
Here $E_{i\bk}=\sqrt{\vare^{2}_{i\bk}+|\De_\bk|^2 }$ are the SC quasiparticle energies where we assumed the same gap function $\De_{i\bk}\equiv \De_\bk$ for each band. Considering only nonmagnetic weak impurity scattering as before the QPI spectrum in the coexisting HO+SC state is then given by
\be
\bl
\Lambda(\bq,\om)
=&
-\frac{1}{\pi} Im \frac{1}{N}\sum_{i\bk}
\bigl[G^i_0(\bk,\om)G^i_0(\bk-\bq,\om)
\\
&
-F^i_0(\bk,\om)F^{i*}_0(\bk-\bq,\om)\bigr],
\el
\ee
leading to the final expression
\be
\bl
&\Lambda(\bq,\om)=
\\
&-\frac{1}{\pi} Im \frac{1}{N}
\sum_{i\bk}
\frac
{(\om+\vare_{i\bk})(\om+\vare_{i\bk-\bq})-\De_\bk\De^{*}_{\bk-\bq}}
{\bigl[(\om)^2-E^{2}_{i\bk}][(\om)^2-E^{2}_{i\bk-\bq} \bigr]}.
\el
\ee
For $\De_\bk =0 $ it reduces to the expression for the normal state with HO in Eq.~(\ref{eq:QPI_HO}).
For the explicit calculation of $\Lambda(\bq,\om)$ in the coexisting HO+SC we need a concrete model for the SC gap function in addition to the HO models defined in Eqs.~(\ref{eq:OP},\ref{eq:HOP}). As mentioned above the chiral d-wave gap function has been proposed from thermal transport and specific heat results. Its explicit form is 
\be
\De_\bk=\De_0\sin\fc k_z\bigl[\sin\fa(k_x+k_y)+\oi\sin\fa(k_x-k_y)\bigr],
\ee
with absolute value,   $|\De_\bk|$, is given by
\be
\bl
\non
&
|\De_\bk|^2=2\De_0^2\sin^2\fc k_z\bigl[\sin^2\fa k_x\cos^2\fa k_y
\\
&\hspace{4cm}
+\cos^2\fa k_x\sin^2\fa k_y\bigr].
\label{eq:chiral-d}
\el
\ee
This gap function has line nodes in the tetragonal plane $k_z=0,\pm 2\pi/c$ which are equivalent
in the folded st BZ. Furthermore it has additional point nodes at locations $k_x=k_y=0$. This may lead to interesting consequences
for the QPI spectrum: For small bias voltage $\omega\ll \Delta_0$ the surfaces $E_{i\bk}=\omega$ have essentially $k_z\simeq 0$ and are unchanged from the non-SC HO state. Consequently the QPI will essentially be only determined by the quasi-2D HO Fermi surface at $k_z=0$ shown in Fig.~\ref{fig:Fig1}. Therefore  the chiral SC gap opening in a way reveals the true HO characteristics in the QPI by reducing it to a 2D situation. However, this also means that one should expect a suppression of any
difference in QPI for $E_-$ and $A_{2+}$ HO in the SC state because in the plane $k_z=0$ they have identical quasiparticle dispersion.

%
%%%%%%%%%%%%%%%%%%%%%%%%%%%%%%%%%%%%%%%%%%%%%%%%%%%%%%%%%%%%%%%%%%%%%
\begin{SCfigure*}
\vspace{0.2cm}
{\includegraphics[width=1.5\linewidth,clip]{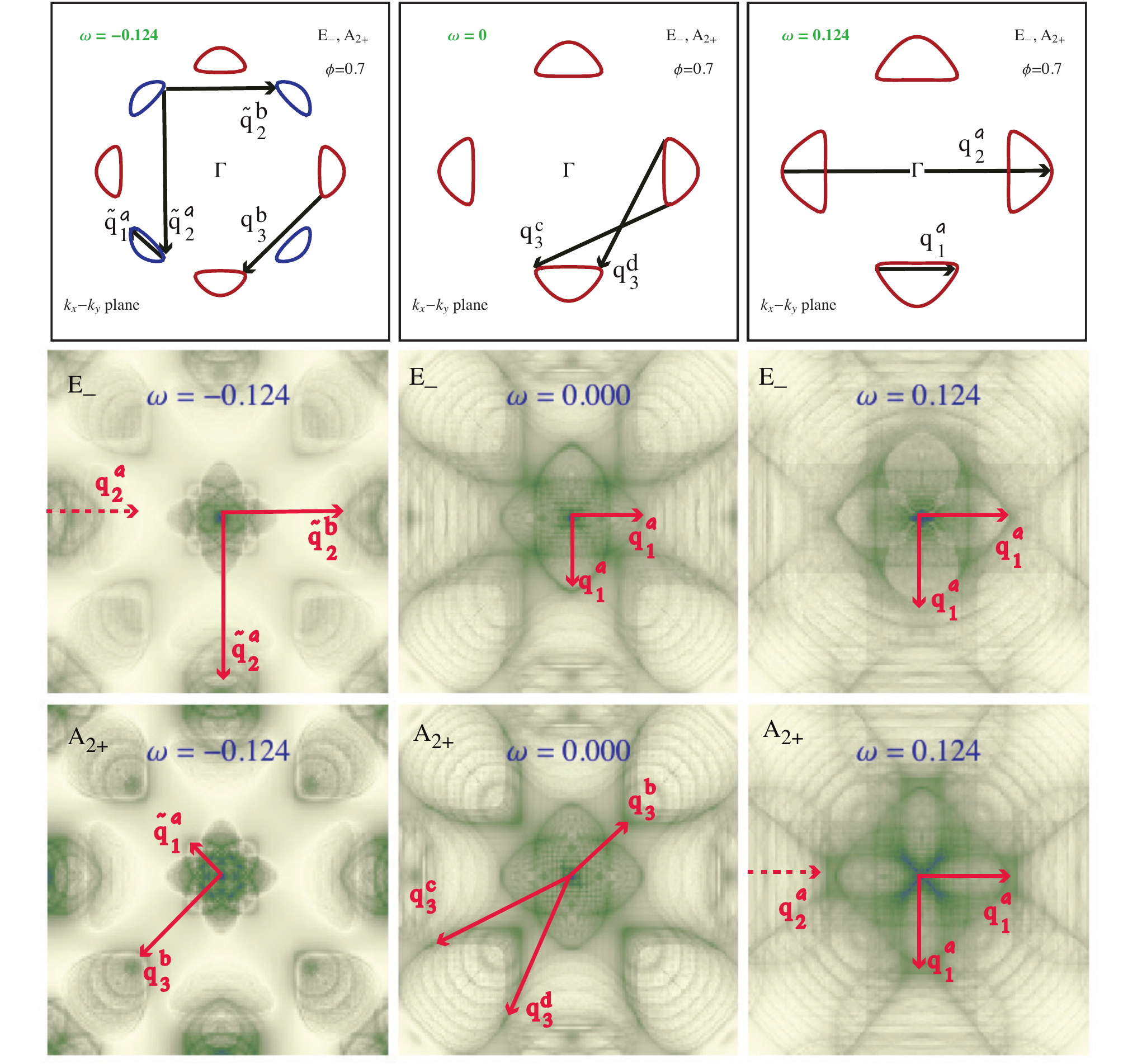}}
\caption{
(Color online) 
Constant energy surfaces ($\epsilon_{i\bk}=\omega$) for $k_z=0$ (top row, identical for both HO) and total QPI (absolute value) spectra ($\omega=0,-0.124t_0, 0.124t_0$) for $E_-$ (center row) and $A_{2+}$ (bottom row) HO parameter.  $C_4$ symmetry is preserved for 
 $A_{2+}$  while a distinct rotational symmetry breaking is still visible in the  integrated QPI for $E_-$. Note that the latter is due to the
 $|k_z|>0$ contributions not depicted in the top row. Dashed arrows denote image folded back into the first BZ.
 Momentum range  is given by $-\pi\leq k_{i}\leq \pi$.
 \vspace{4.5cm}
 }
\label{fig:Fig7}
\end{SCfigure*}
%%%%%%%%%%%%%%%%%%%%%%%%%%%%%%%%%%%%%%%%%%%%%%%%%%%%%%%%%%%%%%%%%%%%%
%

\section{Numerical results and discussion}
\label{sec:numerical}

Using the previous analysis we may now predict all essential QPI spectral properties of the HO and coexisting HO+SC phase. We first specify the numerical parameters. Those for the effective $5f$-bands of the para phase are cited in Appendix \ref{sec:app1}. 
The scale of the quasiparticle bandwidth $W_{qp}\simeq 80$ meV in \URU~ is taken from Ref.~\onlinecite{park:12}. In units of   $t_0=6.66$ meV \cite{rau:12} we have $W_{qp}=12t_0$.

In the HO phase the gap at nesting points is  $\Delta_{HO}=\phi/\sqrt{2}$. From Ref. \onlinecite{aynajian:10} we have $\Delta_{HO}=4.1$ meV or $\phi=5.8$ meV. With the smaller T$_{HO}=16$ K for gap onset \cite{aynajian:10} this leads to a HO BCS ratio 2$\Delta_{HO}/kT_{HO}=5.8$. Furthermore we have $\Delta_{HO}/W_{qp}=5.1\times10^{-2}$. This sizable HO gap value is favorable for clear structure formation in QPI. The corresponding  maximum amplitude of HO is $\phi=7.25\times10^{-2}W_{qp}$ or $\phi= 0.87t_0$.

In the superconducting state we have the average gap value from Ref.~\onlinecite{okazaki:10} with $2\Delta_0/kT_c$=5.6. With $T_c=1.45$ K this means a SC gap amplitude of $\Delta_0=0.35$ meV which is an order of magnitude smaller than the HO gap.
Furthermore we get  $\Delta_{0}/W_{qp}=0.44\times10^{-2}$ or $\Delta_{0}/t_0=0.05$ . To enhance the QPI structures induced by SC state more clearly we will also use a larger value for $\Delta_0$.\\

First we discuss basic characteristics of the para phase model Fermi surface (Fig.~\ref{fig:Fig1}) and band structure (Fig.~\ref{fig:Fig2}a) which was proposed by Rau and Kee \cite{rau:12}. In the unfolded para phase BZ there is an electron sheet around $\Gamma$ and a hole sheet around Z which have comparative sizes and large portions that are nested with $\bQ = (0,0,1)$. In the folded BZ the latter is projected to $\Gamma$, in a corresponding $k_x-k_y$- plane cut (Fig.~\ref{fig:Fig1}b) the nested regions are around the crossing points. Therefore the FS sheets will break up around these crossings in the HO phase. The complementary  $k_x-k_z$ - plane cut is shown in Fig.~\ref{fig:Fig1}c. The band structure in the folded st BZ is shown in Fig.~\ref{fig:Fig2}a and exhibits again the crossing of electron-like and hole-like branches, e.g. along $\Gamma \mbox{X}$.\\

Turning on the HO parameter leads to a repulsion of bands at the crossing point, opening a gap locally (Figs.~\ref{fig:Fig2}b,c). This results in a sharp dip in the DOS at the Fermi level (Fig.~\ref{fig:Fig2}d.) as function of increasing HO strength. Such drastic decrease in the quasiparticle DOS in the HO phase was indeed seen in transport measurements \cite{bel:04,behnia:05}.
The Fermi surface reconstruction in the HO phase is shown in Fig.~\ref{fig:Fig3}a in a 3D representation in the {\it folded} BZ (c.f. Fig.\ref{fig:Fig1}a in the {\it unfolded} BZ). The HO introduces a void in the formerly closed FS body at the Z-points and slices the FS parallel to $k_z$ at the crossing points of the nested sheets.
In the corresponding $k_x-k_y$ - plane cut (Fig.~\ref{fig:Fig3}b) the formerly closed and rounded square-like sheets (Fig.~\ref{fig:Fig1}b) therefore break up into four smaller and four larger petal-like shapes. The smaller ones vanish when the HO parameter $\phi = |\bphi^\bQ|$ or $|\phi^\bQ_z|$ is increased still further. The $k_x-k_z$ - plane cut in Fig.~\ref{fig:Fig3}c in comparison with Fig.~\ref{fig:Fig1}c shows again the vanishing of the FS around the Z-point when the HO gap opens.\\

These main features are similar for both $E_-$ and $A_{2+}$ HO symmetries. However one can identify subtle differences in the reconstructed FS. They are not present for $k_z$=0 cuts because in this case the HO reconstructed dispersions in Eqs.~(\ref{eq:dispE},\ref{eq:dispA2}) are formally equivalent (second of Eq.~(\ref{eq:special})). The difference appears in $k_x-k_y$ - plane cuts for $|k_z| > 0$ as shown in Figs.~\ref{fig:Fig4}a,b. For the $E_-(1,1)$ HO (a) clearly the $C_4$ rotational symmetry of FS sheets is destroyed. The symmetry breaking to $C_2$ would be rotated by $\pi/2$ for the other $E_-(1,\bar{1})$ domain. This asymmetry is absent for the $A_{2+}$ HO (b) where all petals still have the same size, preserving $C_4$ symmetry. Therefore the set of characteristic wave vectors connecting the tips of the petals which should be seen in QPI will be different in the two cases. In Fig.~\ref{fig:Fig4}c we present a zoomed DOS in the HO gap region which shows that the DOS at the Fermi level is strongly reduced when HO develops. This agrees with the experimental observations \cite{bel:04,behnia:05}. \\

Now we discuss the main results of QPI calculations. As mentioned before \URU~ is not an ideal case for the STM-QPI method due to its largely 3D electronic structure. In such a case we proceed in two steps \cite{akbari:11}: First we calculate the QPI spectrum of slices of a given $k_z$ component for wave vectors in the tetragonal plane. This allows one to identify directly the effect of Fermi surface and hidden order on each contribution. For the total QPI spectrum that has to be compared with experiment one must integrate over all slices of different $k_z$, then the question is how much of the characteristic Fermi surface and HO structures in the spectrum survive after the integration and can still be used as a diagnostic of the HO state. In all following figures we show only the absolute value $|\Lambda(\bq,\omega)|$ of the QPI spectrum. \\
%
%%%%%%%%%%%%%%%%%%%%%%%%%%%%%%%%%%%%%%%%%%%%%%%%%%%%%%%%%%%%%%%%%%%%%
\begin{SCfigure*}
%\vspace{0.2cm}
{\includegraphics[width=1.4\linewidth,clip]{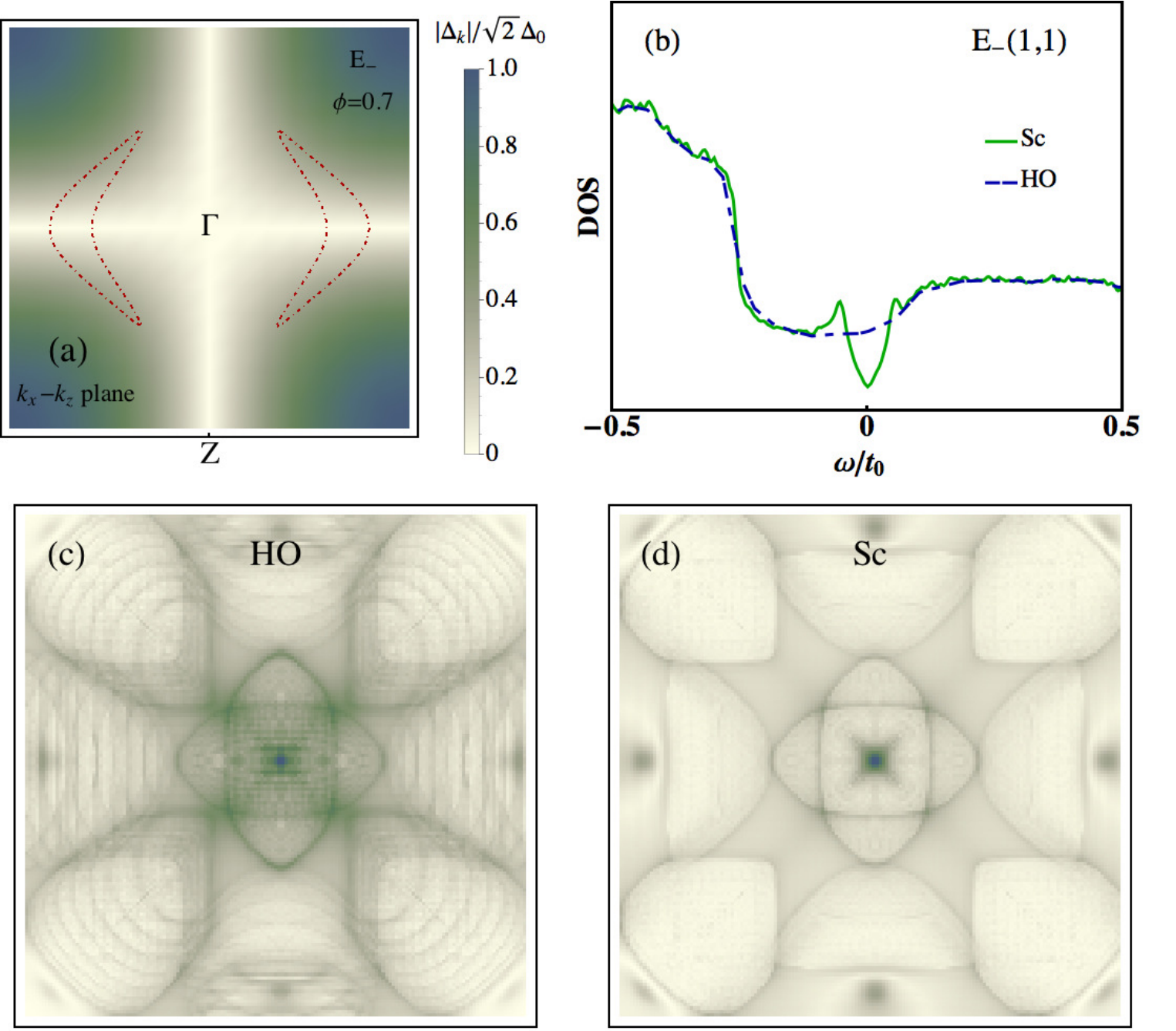}}
\caption{
(Color online) 
(a) Size of the gap function (Eq.~(\ref{eq:chiral-d})) on the HO Fermi surface in $k_x-k_z$ plane. %Scale of $|\Delta_\bk|$ is arbitrary. 
 The dashed red curves show the spectral function at energy $\omega=0$ for hidden ordered phase and blue-thick curves show the quasi-2D spectral function at energy $\omega=0.3\Delta_0$ in HO+SC phase.
 Note the node line $\Delta_\bk=0$ is in $k_z=0$ plane. (b) The DOS shows the evolution of the SC pseudo-gap on top of the larger HO gap. 
Comparison of total QPI  (absolute value) image of (c) $E_-$ HO phase ($\phi=0.7$) (equivalent to Fig.~\ref{fig:Fig7}d) with (d) HO ($E_-$ or $A_{2+}$) +chiral d-wave SC phase ($\phi=0.7,\Delta_0=0.1$). In the HO phase (c) the $k_z$ summation leads to the appearance of $C_4$ symmetry breaking. In the HO+SC phase the gapping of states with $|k_z|>0$ in (a) reduces (d) to a quasi-2D QPI spectrum that contains only the $C_4$ symmetric $k_z=0$ slice and therefore is equivalent for $E_-$ or $A_{2+}$ HO. 
Momentum range is given by $-\pi\leq k_{i}\leq \pi$.
\vspace{0.85cm}
 }
\label{fig:Fig8}
\end{SCfigure*}
%%%%%%%%%%%%%%%%%%%%%%%%%%%%%%%%%%%%%%%%%%%%%%%%%%%%%%%%%%%%%%%%%%%%%
%
In Fig.~\ref{fig:Fig5} we show the FS cut and corresponding QPI spectrum (absolute value) for the $k_z=0$ slice and $\phi=0.7t_0$ which is identical for both HO symmetries.  It is the most important one because the $v^z_{i\bk}=\partial \vare_{i\bk}/\partial k_z $ velocity components vanish for $k_z=0$ (Fig.~\ref{fig:Fig3}a) leading to a large resultant contribution with $k_z\approx 0$ neighbouring slices. The characteristic intra-sheet  
and inter-sheet 
scattering wave vectors of the HO reconstructed Fermi surface are shown in Fig.~\ref{fig:Fig5}a. They should reappear prominently in the calculated  QPI spectrum of Fig.~\ref{fig:Fig5}b. Indeed most of them can be
clearly identified. As a whole the reconstructed HO Fermi surface can  be well recognized in the QPI image in Fig.~\ref{fig:Fig5}b if one keeps in mind that in the latter the characteristic length of  Fermi vectors $k_F$ in Fig.~\ref{fig:Fig5}a  will be mapped to $2k_F$. This means that some of the features like the `petal'-images produced by $\bq^{a-c}_2$ scattering look ,inverted' because they are folded back from the next BZ, leading to effective characteristic wave vectors  $\bq^{a-c}_2 \rightarrow \bq^{a-c}_2-\bK$ (dashed  arrows in  Fig.~\ref{fig:Fig5}b) with \bK~denoting a reciprocal lattice vector.

The $k_z=0$ slice QPI image in Fig.~\ref{fig:Fig5}b  is identical for $E_-$ and $A_{2+}$ HO and has the fourfold $C_4$ symmetry. For $|k_z| > 0$ slices they should become distinct and the symmetry breaking for $E_-$ QPI image should appear corresponding to the Fermi surface cut in Fig.~\ref{fig:Fig4}a. This happens gradually because the symmetry breaking term in Eqs.~(\ref{eq:dispE},\ref{eq:zeta2})  behaves like $\zeta_\bk\sim sin^2\fc k_z$. Thus for $k_z/(\pi/c)\ll1$ the QPI image will be qualitatively as in Fig.~\ref{fig:Fig5}b. For larger $k_z$ it changes rapidly as seen in Fig.~\ref{fig:Fig6}. The dimensions generally shrink because of the reduction
of Fermi surface dimensions obvious from  Fig.~\ref{fig:Fig3}a. But more importantly, while the $A_{2+}$ image in Fig.~\ref{fig:Fig6} retains fourfold symmetry for all $k_z$ the $E_-$ image develops twofold anisotropy for increasing $k_z$. This is already visible
for $k_z=0.3$ since the axis oriented lobes have different widths and the diagonal lobes are not completely symmetric. for even larger $k_z$ the structure of the $E_-$ QPI image changes and develops pronounced twofold symmetry, in contrast to the $A_{2+}$ QPI image which is perfectly $C_4$ symmetric for all slices.\\

The question is now how much of these intricate features in Figs.~\ref{fig:Fig5} and \ref{fig:Fig6} for the individual slices will survive
in the integrated QPI spectrum which can be measured. It is shown in Fig.~\ref{fig:Fig7} for three frequencies (bias voltages) $\omega=0,\pm0.124t_0$ and $E_-$ (second row) and $A_{2+}$ (third row) HO with $\phi=0.7t_0$. Note that the value of the bias voltage is still considerably smaller than the HO gap $\Delta_{HO}=0.62t_0$.
The top row shows the spectral function (equal energy surfaces) in the three cases. While $\omega/t_0=0,0.124$ are similar except for the larger sheet dimensions the case for negative bias $\omega/t_0=-0.124$ is distinct. It can be seen that the diagonal smaller petals (blue) are reintroduced similar to the case for $\phi=0.3t_0$ with $\omega=0$ (Fig.~\ref{fig:Fig3}b). In other words the increasing $\phi$ can be partly compensated by going to negative bias voltage. These introduces new intra-and inter-sheet scattering processes connected with the smaller (blue) petals and labeled by $\tilde{\bq}^a_1$, $\tilde{\bq}^{a,b}_2$ in addition to the $\bq_i^\alpha$ already defined before (Fig.~\ref{fig:Fig5}a). In general the intra-petal scattering is still very clearly visible in the integrated QPI spectra leading to the center ellipsoids. The axis oriented ellipsoids are always present and their size depends on $\omega$, i.e. on the size of the larger (red) petals. On the other hand
the diagonal ellipsoids in QPI spectrum are only there for negative bias voltage because they originate from the smaller (blue) petals as the characteristic wave vectors $\tilde{\bq}_1^a$ clearly indicate. 
The $\Gamma$-centered ellipsoids lead to envelopes that a appear as two nested squares with diagonal (large) and axis (small) orientation (Fig.~\ref{fig:Fig7} for $\omega =0$).

Furthermore the diagonal corner lobes are associated with inter-sheet scattering   ($\bq_3^{b-d}$-type) between the large (red) petals. Generally inter-sheet scattering features are diminished and distorted due to the $k_z$ summation. In fact for $\omega/t_0=0.124$ the diagonal corner lobes have become invisible and only a remnant of the axis-aligned $\bq^a_2$-type inter-pocket scattering remains. On the other hand for  $\omega/t_0=-0.124$ the $\tilde{\bq}_2^{a,b}$-type inter-sheet scattering from the small (blue) petals  is visible in the QPI as new axis aligned lobes. They are, however, superposed to the weak $\bq^a_2$-type scattering between the larger petals.
Most importantly Fig.~\ref{fig:Fig7} (second row) demonstrates that clear but subtle evidence for the $C_4$ symmetry breaking remains in the total QPI for $E_-$. For $\omega=0, -0.124t_0$ clearly the corner lobes from inter-pocket $\bq_3^{b-d}$-type scattering break the $C_4$ symmetry. For $\omega=0.124t_0$ these are no longer visible but the center lobes now have different width for those oriented along $k_x,k_y$ directions. On the other hand the total $A_{2+}$ QPI spectrum keeps full $C_4$ symmetry for all frequencies.

In the experimental QPI spectrum \cite{schmidt:10} the two nested square envelopes of the $\Gamma$-centered ellipsoids from intra-petal scattering are clearly visible where their relative intensity changes with frequency (voltage) . For the lower frequencies the axis and diagonal oriented lobes at larger wave vectors due to inter-petal scattering are also present. However there seems to be no easily recognizable rotational symmetry breaking present. As mentioned the latter may ony appear in single domain samples of $E_-$ type HO.

Finally we discuss the influence of the proposed chiral superconducting order within HO state on the QPI spectrum. In  Fig.~\ref{fig:Fig8}a the spectral function in HO phase (dashed red line) is superposed to the SC gap contour plot in $k_x-k_z$ plane. Clearly the node line $k_z=0$ crosses the FS sheet while the node points miss it. This means in the SC phase we obtain an effectively 2D slice around $k_z=0$ for the spectral function (blue full line).
The embedding of SC into the HO phase is demonstrated by the two superposed gaps in the DOS of Fig.~\ref{fig:Fig8}b. For QPI we assume the case $\omega<\Delta_0\ll\Delta_{HO}$. Due to the $k_z=0$ a node line of the SC the gap opens up only for $|k_z|>0$ slices and strongly reduces their contribution to the total QPI. Then the latter becomes {\it effectively 2D} in the SC phase despite the 3D Fermi surface. Therefore the chiral d-wave SC state should be favorable to unveil the QPI structures of the HO phase with more clarity as demonstrated by the comparison of Figs.~\ref{fig:Fig8}c,d. Indeed the full QPI of Fig.~\ref{fig:Fig8}d in the SC+HO state is practically identical to the partial QPI of $k_z=0$ slice in the HO phase (Fig.~\ref{fig:Fig5}b) without SC gap. This is obvious from the quasi-2D shape of the (blue) spectral function in Figs.~\ref{fig:Fig8}a.  Although the $k_z$=0 node line of the gap is favorable for enhancing the QPI features due to HO there is a drawback: Because $|k_z|>0$ contributions will be suppressed the $C_4$ symmetry breaking in the $E_-$ phase which they cause will also be suppressed. Therefore in the HO+SC phase the QPI pattern of  $E_-$ and $A_{2+}$ will be indistinguishable for $\omega <\Delta_0\ll\Delta_{HO}$.

\section{Summary and Conclusion}
\label{sec:summary}

In this work we have performed an analysis of the consequences and signature of hidden order and chiral d-wave superconductivity  in the STM quasiparticle interference of \URU. This work was motivated on one hand by already existing experiments and on the other hand by the continuing debate on the proper symmetry of the hidden order. Our calculations are based on a simplified effective model of Ref.~\onlinecite{rau:12} that reproduces the main nesting Fermi surface sheets of \URU~thought to be responsible for the multipolar HO. In particular we studied the two most frequently discussed order parameter symmetries, the $A_{2+}$ (rank 4) hexadecapole  and $E_-$ (rank 5) dotriacontapole. Existing experimental evidence (Sec.~\ref{sec:multipolar}) favors the latter. Their main difference is the presence of $C_4$ fourfold to twofold symmetry breaking in the latter which is absent in the former. These order parameters may be described by electron-hole pairing with nesting momentum \bQ~ in different total angular momentum states.

The calculation of quasiparticle bands clearly shows the breakup of large electron hole pockets of the disordered phase into smaller pockets (petals) in the HO state. While the reconstructed Fermi surface sheets are equivalent in the tetragonal plane for both symmetries, they strongly differ away from it $(|k_z|>0)$ through the presence or absence of fourfold symmetry. The symmetry breaking for $E_-$ is directly related to the presence of inter-orbital hopping terms. In both HO cases a deep gap in the DOS evolves in accordance with experimental observation.

The quasiparticle interference spectrum was calculated in Born approximation using the four reconstructed bands. Due to the pronounced 3D character of the Fermi surface in disordered as well as HO phases the QPI calculations has to be performed for 2D slices of constant $k_z$ followed by a summation. The result shows the main features of the FS reconstruction by HO can still be seen in the total QPI spectrum at various bias voltages. The most prominent features result from intra-sheet scattering while the structures due to inter-petal scattering are more diffuse and depend on the bias voltage size. The presence of the QPI center ellipsoids and corner or edge lobe structures and partly its frequency dependence is qualitatively similar to the experimental results \cite{schmidt:10}. The center ellipsoids may also may also be interpreted as nested axis- and diagonal- oriented squares.

The calculation has also shown that there are subtle distinctions between the HO symmetries, in particular the clear fourfold symmetry breaking of QPI pattern in the $E_-$ phase in contrast to $A_{2+}$ (Fig.~\ref{fig:Fig7}). This is, however, observable only for scanning a single domain of the $E_-$ phase. Averaging over domains would restore the fourfold symmetry also in the $E_-$ HO phase. This seems to be the case in presently existing experiments \cite{schmidt:10}. The importance of having single domain samples for observing the fourfold symmetry breaking was already emphasized in torque experiments \cite{okazaki:11}  and their theoretical interpretation \cite{thalmeier:11}. If single domain samples can be realized in these QPI experiments they can give additional evidence for the HO symmetry.

The influence of the frequently discussed chiral d-wave SC order embedded in the HO phase has been investigated. Due to its node line in the basal plane this order parameter effectively leads to a reduction of total 3D QPI to an essentially 2D spectrum with improved contrast but at the same time it suppresses any symmetry distinction between $E_-$ and $A_{2+}$ hidden order.

\appendix
\section{Kinetic energy coefficients and para phase band structure}
\label{sec:app1}

Here we describe the effective 5f- two band model for \URU~ that is adopted from Ref.~\onlinecite{rau:12}. The kinetic terms in 
Eq.~(\ref{eq:bands}) are defined by the intra-orbital energies ($\al =1,2$ is the orbital or band index):
\be
\bl
A_{\al\bk}=&A_{\al\bk}^z+A_{\al\bk}^\perp +\frac{1}{2}sign(\al)\Delta_{12} ,
%\non
\\
A_{\al\bk}^z=&8t_\al\cos\fa k_x\cos\fa k_y\cos\fc k_z,
%\non
\\
A_{\al\bk}^\perp=&
2t'_\al(\cos ak_x +\cos ak_y) 
%\non\\&
+4t''_\al\cos ak_x\cos ak_y-\epsilon_0,
%\non
\el
\ee
and inter-orbital hopping energy
\be
\bl
D_\bk=
&
D'_\bk+iD''_\bk
%\non
\\
 =&t_{12}\bigl[\sin\fa(k_x+k_y)
-
\oi
\sin\fa(k_x-k_y)\bigr]\sin\fc k_z,
%\non
\\
D'_\bk=&4t_{12}\sin\fa(k_x+k_y)\sin\fc k_z,
%\non
\\
D''_\bk=&-4t_{12}\sin\fa(k_x-k_y)\sin\fc k_z.
\label{eq:kinetic}
\el
\ee
To reproduce a realistic Fermi surface model with nesting electron- and hole- like pockets 
around the $\Gamma$ and $Z$ points of the bcc Brillouin zone we use the following parameters \cite{rau:12}:
The orbital energy splitting is  $\Delta_{12}=3.5$ or $\Delta\equiv 0.5 \Delta_{12}=1.75$.
The nearest neighbor hopping is $t_1=t_2\equiv t =-0.3$, this means orbital-independent $A_{\al\bk}^z=A_{\bk}^z$. Furthermore hopping elements to next and second nearest neighbors are given by 
$t'_1=-0.87$, $t'_2=0.0, t''_1=0.375, t''_2=0.25$, respectively and the average orbital energy is $-\epsilon_0 = 0.5$. The inter-orbital hopping is taken as $|t_{12}|=0.7$.
All energies are given here in terms of the unit $t_0$. Since the total effective band width (Fig.\ref{fig:Fig2}a) is $W_{qp}\simeq 12t_0$ and $W_{qp}=80$ meV from tunneling results \cite{park:12} this means $t_0=6.66$ meV.

For the computation of quasiparticle bands in the HO phase it is also useful to introduce the following (anti-) symetrized
quantities:
\be
\bl
A_\bk^\perp=&\frac{1}{2}(A_{1\bk}^\perp+A_{2\bk}^\perp)
\\
=&2t'(\cos ak_x +\cos ak_y) +4t''\cos ak_x\cos ak_y -\epsilon_0,
\\
\De_\bk^\perp=&\Delta+\frac{1}{2}(A_{1\bk}^\perp-A_{2\bk}^\perp)
\\
=&
\Delta+2\de'(\cos ak_x +\cos ak_y) +4\de''\cos ak_x\cos ak_y.
\label{eq:kinsymm}
\el
\ee
Here we defined $t'=\frac{1}{2}(t'_1+t'_2)$,  $t''=\frac{1}{2}(t''_1+t''_2)$ and  
$\de'=\frac{1}{2}(t'_1-t'_2)$,  $\de''=\frac{1}{2}(t''_1-t''_2)$.\\
The auxiliary functions above have the following symmetry under translation by the ordering vector \bQ:
$A_{\al\bk+\bQ}^\perp=A_{\al\bk}^\perp$ implying also $A_{\bk+\bQ}^\perp=A_{\bk}^\perp$ and  
$\De_{\bk+\bQ}^\perp=\De_{\bk}^\perp$. On the  other hand $A_{\al\bk+\bQ}^z=-A_{\al\bk}^z$ and
$D_{\bk+\bQ}=-D_\bk$.\\

Finally we give the relations of the $A_{1\bk}$, $A_{2\bk}$ to the symmetrized coefficients $A^\perp_\bk$, $\Delta^\perp_\bk$ introducing 
$\tau=\pm$ as the new band index  connected with the downfolding of the paramagnetic bct to the st BZ of the HO phase and the associated symmetry $A^z_{\bk+\bQ}=-A^z_\bk$. We obtain\\
\be
\bl
A^\bk_{\tau +}=\tau A^z_\bk + A^\perp_{\bk}+\Delta^\perp_\bk=
\left\{
\begin{array}{cc}
A_{1\bk}  & \tau=+
\\
A_{1\bk+\bQ} & \tau=-
\end{array}
\right\},
\\\\
A^\bk_{\tau -}=\tau A^z_\bk + A^\perp_{\bk}-\Delta^\perp_\bk=
\left\{
\begin{array}{cc}
A_{2\bk}  & \tau=+
\\
A_{2\bk+\bQ} & \tau=-
\end{array}
\right\}.
\label{eq:auxiliary}
\el
\ee
This leads to the identities
\be
\bl[l]
&\frac{1}{2}(A_{1\bk}+A_{2\bk}) =A^z_\bk+A^\perp_\bk, %;\;\;\;\;\; 
\\
&\frac{1}{2}(A_{1\bk}-A_{2\bk})=\De^\perp_\bk,
\el
\ee
and 
\be
\bl
&\frac{1}{2}(A_{1\bk+\bQ}+A_{2\bk+\bQ}) =-A^z_\bk+A^\perp_\bk,
\\
&\frac{1}{2}(A_{1\bk+\bQ}-A_{2\bk+\bQ})=\De^\perp_\bk .
\label{eq:downfold}
\el
\ee
The para phase ($|\bphi|=|\phi_z|=0$) band structure is given by the first of Eq.~(\ref{eq:special}). 
Using Eq.~(\ref{eq:downfold}) these four bands may also be expressed as
\be
\bl
\vare^\pm_1(\bk)=&
\frac{1}{2}(A_{1\bk}+A_{2\bk}) \pm
\sqrt{\frac{1}{4}(A_{1\bk}-A_{2\bk})^2+|D_\bk|^2},
\\
\\
\vare^\pm_2(\bk)=&
\frac{1}{2}(A_{1\bk +\bQ}+A_{2\bk +\bQ})
\\&
\pm
\sqrt{\frac{1}{4}(A_{1\bk +\bQ}-A_{2\bk +\bQ})^2+|D_\bk|^2},
\label{eq:rau-equivalence}
\el
\ee
which reflects directly that the four bands are obtained by downfolding of two bands into the st BZ of the ordered phase. 

\bibliography{References}

\begin{thebibliography}{35}%
\makeatletter
\providecommand \@ifxundefined [1]{%
 \@ifx{#1\undefined}
}%
\providecommand \@ifnum [1]{%
 \ifnum #1\expandafter \@firstoftwo
 \else \expandafter \@secondoftwo
 \fi
}%
\providecommand \@ifx [1]{%
 \ifx #1\expandafter \@firstoftwo
 \else \expandafter \@secondoftwo
 \fi
}%
\providecommand \natexlab [1]{#1}%
\providecommand \enquote  [1]{``#1''}%
\providecommand \bibnamefont  [1]{#1}%
\providecommand \bibfnamefont [1]{#1}%
\providecommand \citenamefont [1]{#1}%
\providecommand \href@noop [0]{\@secondoftwo}%
\providecommand \href [0]{\begingroup \@sanitize@url \@href}%
\providecommand \@href[1]{\@@startlink{#1}\@@href}%
\providecommand \@@href[1]{\endgroup#1\@@endlink}%
\providecommand \@sanitize@url [0]{\catcode `\\12\catcode `\$12\catcode
  `\&12\catcode `\#12\catcode `\^12\catcode `\_12\catcode `\%12\relax}%
\providecommand \@@startlink[1]{}%
\providecommand \@@endlink[0]{}%
\providecommand \url  [0]{\begingroup\@sanitize@url \@url }%
\providecommand \@url [1]{\endgroup\@href {#1}{\urlprefix }}%
\providecommand \urlprefix  [0]{URL }%
\providecommand \Eprint [0]{\href }%
\providecommand \doibase [0]{http://dx.doi.org/}%
\providecommand \selectlanguage [0]{\@gobble}%
\providecommand \bibinfo  [0]{\@secondoftwo}%
\providecommand \bibfield  [0]{\@secondoftwo}%
\providecommand \translation [1]{[#1]}%
\providecommand \BibitemOpen [0]{}%
\providecommand \bibitemStop [0]{}%
\providecommand \bibitemNoStop [0]{.\EOS\space}%
\providecommand \EOS [0]{\spacefactor3000\relax}%
\providecommand \BibitemShut  [1]{\csname bibitem#1\endcsname}%
\let\auto@bib@innerbib\@empty
%</preamble>
\bibitem [{\citenamefont {Mydosh}\ and\ \citenamefont
  {Oppeneer}(2011)}]{mydosh:11}%
  \BibitemOpen
  \bibfield  {author} {\bibinfo {author} {\bibfnamefont {J.~A.}\ \bibnamefont
  {Mydosh}}\ and\ \bibinfo {author} {\bibfnamefont {P.~M.}\ \bibnamefont
  {Oppeneer}},\ }\href@noop {} {\bibfield  {journal} {\bibinfo  {journal} {Rev.
  Mod. Phys.}\ }\textbf {\bibinfo {volume} {83}},\ \bibinfo {pages} {1301}
  (\bibinfo {year} {2011})}\BibitemShut {NoStop}%
\bibitem [{\citenamefont {Santini}\ and\ \citenamefont
  {Amoretti}(1994)}]{santini:94}%
  \BibitemOpen
  \bibfield  {author} {\bibinfo {author} {\bibfnamefont {P.}~\bibnamefont
  {Santini}}\ and\ \bibinfo {author} {\bibfnamefont {G.}~\bibnamefont
  {Amoretti}},\ }\href@noop {} {\bibfield  {journal} {\bibinfo  {journal}
  {Phys. Rev. Lett.}\ }\textbf {\bibinfo {volume} {73}},\ \bibinfo {pages}
  {1027} (\bibinfo {year} {1994})}\BibitemShut {NoStop}%
\bibitem [{\citenamefont {Meng}\ \emph {et~al.}(2013)\citenamefont {Meng},
  \citenamefont {Oppeneer}, \citenamefont {Mydosh}, \citenamefont
  {Riseborough}, \citenamefont {Gofryk}, \citenamefont {Joyce}, \citenamefont
  {Bauer}, \citenamefont {Li},\ and\ \citenamefont {Durakiewicz}}]{meng:13}%
  \BibitemOpen
  \bibfield  {author} {\bibinfo {author} {\bibfnamefont {J.-Q.}\ \bibnamefont
  {Meng}}, \bibinfo {author} {\bibfnamefont {P.~M.}\ \bibnamefont {Oppeneer}},
  \bibinfo {author} {\bibfnamefont {J.~A.}\ \bibnamefont {Mydosh}}, \bibinfo
  {author} {\bibfnamefont {P.~S.}\ \bibnamefont {Riseborough}}, \bibinfo
  {author} {\bibfnamefont {K.}~\bibnamefont {Gofryk}}, \bibinfo {author}
  {\bibfnamefont {J.~J.}\ \bibnamefont {Joyce}}, \bibinfo {author}
  {\bibfnamefont {E.~D.}\ \bibnamefont {Bauer}}, \bibinfo {author}
  {\bibfnamefont {Y.}~\bibnamefont {Li}}, \ and\ \bibinfo {author}
  {\bibfnamefont {T.}~\bibnamefont {Durakiewicz}},\ }\href@noop {} {\bibfield
  {journal} {\bibinfo  {journal} {Phys. Rev. Lett.}\ }\textbf {\bibinfo
  {volume} {111}},\ \bibinfo {pages} {127002} (\bibinfo {year}
  {2013})}\BibitemShut {NoStop}%
\bibitem [{\citenamefont {Oppeneer}\ \emph {et~al.}(2010)\citenamefont
  {Oppeneer}, \citenamefont {Rusz}, \citenamefont {Elgazzar}, \citenamefont
  {Suzuki}, \citenamefont {Durakiewicz},\ and\ \citenamefont
  {Mydosh}}]{oppeneer:10}%
  \BibitemOpen
  \bibfield  {author} {\bibinfo {author} {\bibfnamefont {P.~M.}\ \bibnamefont
  {Oppeneer}}, \bibinfo {author} {\bibfnamefont {J.}~\bibnamefont {Rusz}},
  \bibinfo {author} {\bibfnamefont {S.}~\bibnamefont {Elgazzar}}, \bibinfo
  {author} {\bibfnamefont {M.-T.}\ \bibnamefont {Suzuki}}, \bibinfo {author}
  {\bibfnamefont {T.}~\bibnamefont {Durakiewicz}}, \ and\ \bibinfo {author}
  {\bibfnamefont {J.~A.}\ \bibnamefont {Mydosh}},\ }\href@noop {} {\bibfield
  {journal} {\bibinfo  {journal} {Phys. Rev. B}\ }\textbf {\bibinfo {volume}
  {82}},\ \bibinfo {pages} {205103} (\bibinfo {year} {2010})}\BibitemShut
  {NoStop}%
\bibitem [{\citenamefont {Ikeda}\ \emph {et~al.}(2012)\citenamefont {Ikeda},
  \citenamefont {Suzuki}, \citenamefont {Arita}, \citenamefont {Takimoto},
  \citenamefont {Shibauchi},\ and\ \citenamefont {Matsuda}}]{ikeda:12}%
  \BibitemOpen
  \bibfield  {author} {\bibinfo {author} {\bibfnamefont {H.}~\bibnamefont
  {Ikeda}}, \bibinfo {author} {\bibfnamefont {M.-T.}\ \bibnamefont {Suzuki}},
  \bibinfo {author} {\bibfnamefont {R.}~\bibnamefont {Arita}}, \bibinfo
  {author} {\bibfnamefont {T.}~\bibnamefont {Takimoto}}, \bibinfo {author}
  {\bibfnamefont {T.}~\bibnamefont {Shibauchi}}, \ and\ \bibinfo {author}
  {\bibfnamefont {Y.}~\bibnamefont {Matsuda}},\ }\href@noop {} {\bibfield
  {journal} {\bibinfo  {journal} {Nature Physics}\ }\textbf {\bibinfo {volume}
  {8}},\ \bibinfo {pages} {528} (\bibinfo {year} {2012})}\BibitemShut {NoStop}%
\bibitem [{\citenamefont {Kasahara}\ \emph {et~al.}(2007)\citenamefont
  {Kasahara}, \citenamefont {Iwasawa}, \citenamefont {Shishido}, \citenamefont
  {Shibauchi}, \citenamefont {Behnia}, \citenamefont {Haga}, \citenamefont
  {Matsuda}, \citenamefont {Onuki}, \citenamefont {Sigrist},\ and\
  \citenamefont {Matsuda}}]{kasahara:07}%
  \BibitemOpen
  \bibfield  {author} {\bibinfo {author} {\bibfnamefont {Y.}~\bibnamefont
  {Kasahara}}, \bibinfo {author} {\bibfnamefont {T.}~\bibnamefont {Iwasawa}},
  \bibinfo {author} {\bibfnamefont {H.}~\bibnamefont {Shishido}}, \bibinfo
  {author} {\bibfnamefont {T.}~\bibnamefont {Shibauchi}}, \bibinfo {author}
  {\bibfnamefont {K.}~\bibnamefont {Behnia}}, \bibinfo {author} {\bibfnamefont
  {Y.}~\bibnamefont {Haga}}, \bibinfo {author} {\bibfnamefont {T.~D.}\
  \bibnamefont {Matsuda}}, \bibinfo {author} {\bibfnamefont {Y.}~\bibnamefont
  {Onuki}}, \bibinfo {author} {\bibfnamefont {M.}~\bibnamefont {Sigrist}}, \
  and\ \bibinfo {author} {\bibfnamefont {Y.}~\bibnamefont {Matsuda}},\
  }\href@noop {} {\bibfield  {journal} {\bibinfo  {journal} {Phys. Rev. Lett.}\
  }\textbf {\bibinfo {volume} {99}},\ \bibinfo {pages} {116402} (\bibinfo
  {year} {2007})}\BibitemShut {NoStop}%
\bibitem [{\citenamefont {Akbari}\ \emph {et~al.}(2011)\citenamefont {Akbari},
  \citenamefont {Thalmeier},\ and\ \citenamefont {Eremin}}]{akbari:11}%
  \BibitemOpen
  \bibfield  {author} {\bibinfo {author} {\bibfnamefont {A.}~\bibnamefont
  {Akbari}}, \bibinfo {author} {\bibfnamefont {P.}~\bibnamefont {Thalmeier}}, \
  and\ \bibinfo {author} {\bibfnamefont {I.}~\bibnamefont {Eremin}},\
  }\href@noop {} {\bibfield  {journal} {\bibinfo  {journal} {Phys. Rev. B}\
  }\textbf {\bibinfo {volume} {84}},\ \bibinfo {pages} {134505} (\bibinfo
  {year} {2011})}\BibitemShut {NoStop}%
\bibitem [{\citenamefont {Allan}\ \emph {et~al.}(2013)\citenamefont {Allan},
  \citenamefont {Massee}, \citenamefont {Morr}, \citenamefont {Dyke},
  \citenamefont {Rost}, \citenamefont {Mackenzie}, \citenamefont {Petrovic},\
  and\ \citenamefont {Davis}}]{allan:13}%
  \BibitemOpen
  \bibfield  {author} {\bibinfo {author} {\bibfnamefont {M.~P.}\ \bibnamefont
  {Allan}}, \bibinfo {author} {\bibfnamefont {F.}~\bibnamefont {Massee}},
  \bibinfo {author} {\bibfnamefont {D.~K.}\ \bibnamefont {Morr}}, \bibinfo
  {author} {\bibfnamefont {J.~V.}\ \bibnamefont {Dyke}}, \bibinfo {author}
  {\bibfnamefont {A.~W.}\ \bibnamefont {Rost}}, \bibinfo {author}
  {\bibfnamefont {A.~P.}\ \bibnamefont {Mackenzie}}, \bibinfo {author}
  {\bibfnamefont {C.}~\bibnamefont {Petrovic}}, \ and\ \bibinfo {author}
  {\bibfnamefont {J.~C.}\ \bibnamefont {Davis}},\ }\href@noop {} {\bibfield
  {journal} {\bibinfo  {journal} {Nature Physics}\ }\textbf {\bibinfo {volume}
  {9}},\ \bibinfo {pages} {468} (\bibinfo {year} {2013})}\BibitemShut {NoStop}%
\bibitem [{\citenamefont {Zhou}\ \emph {et~al.}(2013)\citenamefont {Zhou},
  \citenamefont {Misra}, \citenamefont {da~Silva~Neto}, \citenamefont
  {Aynajian}, \citenamefont {Baumbach}, \citenamefont {Bauer},\ and\
  \citenamefont {Yazdani}}]{zhou:13}%
  \BibitemOpen
  \bibfield  {author} {\bibinfo {author} {\bibfnamefont {B.~B.}\ \bibnamefont
  {Zhou}}, \bibinfo {author} {\bibfnamefont {S.}~\bibnamefont {Misra}},
  \bibinfo {author} {\bibfnamefont {E.~H.}\ \bibnamefont {da~Silva~Neto}},
  \bibinfo {author} {\bibfnamefont {P.}~\bibnamefont {Aynajian}}, \bibinfo
  {author} {\bibfnamefont {R.~E.}\ \bibnamefont {Baumbach}}, \bibinfo {author}
  {\bibfnamefont {J.~D. T. E.~D.}\ \bibnamefont {Bauer}}, \ and\ \bibinfo
  {author} {\bibfnamefont {A.}~\bibnamefont {Yazdani}},\ }\href@noop {}
  {\bibfield  {journal} {\bibinfo  {journal} {Nature Physics}\ }\textbf
  {\bibinfo {volume} {9}},\ \bibinfo {pages} {474} (\bibinfo {year}
  {2013})}\BibitemShut {NoStop}%
\bibitem [{\citenamefont {Schmidt}\ \emph {et~al.}(2010)\citenamefont
  {Schmidt}, \citenamefont {Hamidian}, \citenamefont {Wahl}, \citenamefont
  {Meier}, \citenamefont {Balatsky}, \citenamefont {Garrett}, \citenamefont
  {Williams}, \citenamefont {Luke},\ and\ \citenamefont {Davis}}]{schmidt:10}%
  \BibitemOpen
  \bibfield  {author} {\bibinfo {author} {\bibfnamefont {A.~R.}\ \bibnamefont
  {Schmidt}}, \bibinfo {author} {\bibfnamefont {M.~H.}\ \bibnamefont
  {Hamidian}}, \bibinfo {author} {\bibfnamefont {P.}~\bibnamefont {Wahl}},
  \bibinfo {author} {\bibfnamefont {F.}~\bibnamefont {Meier}}, \bibinfo
  {author} {\bibfnamefont {A.~V.}\ \bibnamefont {Balatsky}}, \bibinfo {author}
  {\bibfnamefont {J.~D.}\ \bibnamefont {Garrett}}, \bibinfo {author}
  {\bibfnamefont {T.~J.}\ \bibnamefont {Williams}}, \bibinfo {author}
  {\bibfnamefont {G.~M.}\ \bibnamefont {Luke}}, \ and\ \bibinfo {author}
  {\bibfnamefont {J.~C.}\ \bibnamefont {Davis}},\ }\href@noop {} {\bibfield
  {journal} {\bibinfo  {journal} {Nature}\ }\textbf {\bibinfo {volume} {465}},\
  \bibinfo {pages} {570} (\bibinfo {year} {2010})}\BibitemShut {NoStop}%
\bibitem [{\citenamefont {Haule}\ and\ \citenamefont
  {Kotliar}(2009)}]{haule:09}%
  \BibitemOpen
  \bibfield  {author} {\bibinfo {author} {\bibfnamefont {K.}~\bibnamefont
  {Haule}}\ and\ \bibinfo {author} {\bibfnamefont {G.}~\bibnamefont
  {Kotliar}},\ }\href@noop {} {\bibfield  {journal} {\bibinfo  {journal}
  {Nature Physics}\ }\textbf {\bibinfo {volume} {5}},\ \bibinfo {pages} {796}
  (\bibinfo {year} {2009})}\BibitemShut {NoStop}%
\bibitem [{\citenamefont {Rau}\ and\ \citenamefont {Kee}(2012)}]{rau:12}%
  \BibitemOpen
  \bibfield  {author} {\bibinfo {author} {\bibfnamefont {J.~G.}\ \bibnamefont
  {Rau}}\ and\ \bibinfo {author} {\bibfnamefont {H.-Y.}\ \bibnamefont {Kee}},\
  }\href@noop {} {\bibfield  {journal} {\bibinfo  {journal} {Phys. Rev. B}\
  }\textbf {\bibinfo {volume} {85}},\ \bibinfo {pages} {245112} (\bibinfo
  {year} {2012})}\BibitemShut {NoStop}%
\bibitem [{\citenamefont {Takimoto}\ and\ \citenamefont
  {Thalmeier}(2008)}]{takimoto:08}%
  \BibitemOpen
  \bibfield  {author} {\bibinfo {author} {\bibfnamefont {T.}~\bibnamefont
  {Takimoto}}\ and\ \bibinfo {author} {\bibfnamefont {P.}~\bibnamefont
  {Thalmeier}},\ }\href@noop {} {\bibfield  {journal} {\bibinfo  {journal}
  {Phys. Rev. B}\ }\textbf {\bibinfo {volume} {77}},\ \bibinfo {pages} {045105}
  (\bibinfo {year} {2008})}\BibitemShut {NoStop}%
\bibitem [{\citenamefont {Thalmeier}\ and\ \citenamefont
  {Takimoto}(2011)}]{thalmeier:11}%
  \BibitemOpen
  \bibfield  {author} {\bibinfo {author} {\bibfnamefont {P.}~\bibnamefont
  {Thalmeier}}\ and\ \bibinfo {author} {\bibfnamefont {T.}~\bibnamefont
  {Takimoto}},\ }\href@noop {} {\bibfield  {journal} {\bibinfo  {journal}
  {Phys. Rev. B}\ }\textbf {\bibinfo {volume} {83}},\ \bibinfo {pages} {165110}
  (\bibinfo {year} {2011})}\BibitemShut {NoStop}%
\bibitem [{\citenamefont {Thalmeier}\ \emph {et~al.}(2013)\citenamefont
  {Thalmeier}, \citenamefont {Takimoto},\ and\ \citenamefont
  {Ikeda}}]{thalmeier:13}%
  \BibitemOpen
  \bibfield  {author} {\bibinfo {author} {\bibfnamefont {P.}~\bibnamefont
  {Thalmeier}}, \bibinfo {author} {\bibfnamefont {T.}~\bibnamefont {Takimoto}},
  \ and\ \bibinfo {author} {\bibfnamefont {H.}~\bibnamefont {Ikeda}},\
  }\href@noop {} {\bibfield  {journal} {\bibinfo  {journal} {Phil. Mag.}\ ,\
  \bibinfo {pages} {DOI: 10.1080/14786435.2013.861615}} (\bibinfo {year}
  {2013})}\BibitemShut {NoStop}%
\bibitem [{\citenamefont {Yuan}\ \emph {et~al.}(2012)\citenamefont {Yuan},
  \citenamefont {Figgins},\ and\ \citenamefont {Morr}}]{yuan:12}%
  \BibitemOpen
  \bibfield  {author} {\bibinfo {author} {\bibfnamefont {T.}~\bibnamefont
  {Yuan}}, \bibinfo {author} {\bibfnamefont {J.}~\bibnamefont {Figgins}}, \
  and\ \bibinfo {author} {\bibfnamefont {D.~K.}\ \bibnamefont {Morr}},\
  }\href@noop {} {\bibfield  {journal} {\bibinfo  {journal} {Phys. Rev. B}\
  }\textbf {\bibinfo {volume} {86}},\ \bibinfo {pages} {035129} (\bibinfo
  {year} {2012})}\BibitemShut {NoStop}%
\bibitem [{\citenamefont {Shibauchi}\ \emph {et~al.}(2014)\citenamefont
  {Shibauchi}, \citenamefont {Ikeda},\ and\ \citenamefont
  {Matsuda}}]{shibauchi:14}%
  \BibitemOpen
  \bibfield  {author} {\bibinfo {author} {\bibfnamefont {T.}~\bibnamefont
  {Shibauchi}}, \bibinfo {author} {\bibfnamefont {H.}~\bibnamefont {Ikeda}}, \
  and\ \bibinfo {author} {\bibfnamefont {Y.}~\bibnamefont {Matsuda}},\
  }\href@noop {} {\bibfield  {journal} {\bibinfo  {journal} {Phil. Mag.}\ ,\
  \bibinfo {pages} {DOI: 10.1080/14786435.2014.887861}} (\bibinfo {year}
  {2014})}\BibitemShut {NoStop}%
\bibitem [{\citenamefont {Yoshida}\ \emph {et~al.}(2013)\citenamefont
  {Yoshida}, \citenamefont {Tsubota}, \citenamefont {Ishiga}, \citenamefont
  {Sunagawa}, \citenamefont {Sonoyama}, \citenamefont {Aoki}, \citenamefont
  {Flouquet}, \citenamefont {Wakita}, \citenamefont {Muraoka},\ and\
  \citenamefont {Yokoya}}]{yoshida:13}%
  \BibitemOpen
  \bibfield  {author} {\bibinfo {author} {\bibfnamefont {R.}~\bibnamefont
  {Yoshida}}, \bibinfo {author} {\bibfnamefont {K.}~\bibnamefont {Tsubota}},
  \bibinfo {author} {\bibfnamefont {T.}~\bibnamefont {Ishiga}}, \bibinfo
  {author} {\bibfnamefont {M.}~\bibnamefont {Sunagawa}}, \bibinfo {author}
  {\bibfnamefont {J.}~\bibnamefont {Sonoyama}}, \bibinfo {author}
  {\bibfnamefont {D.}~\bibnamefont {Aoki}}, \bibinfo {author} {\bibfnamefont
  {J.}~\bibnamefont {Flouquet}}, \bibinfo {author} {\bibfnamefont
  {T.}~\bibnamefont {Wakita}}, \bibinfo {author} {\bibfnamefont
  {Y.}~\bibnamefont {Muraoka}}, \ and\ \bibinfo {author} {\bibfnamefont
  {T.}~\bibnamefont {Yokoya}},\ }\href@noop {} {\bibfield  {journal} {\bibinfo
  {journal} {Sci. Rep.}\ }\textbf {\bibinfo {volume} {3}},\ \bibinfo {pages}
  {2750} (\bibinfo {year} {2013})}\BibitemShut {NoStop}%
\bibitem [{\citenamefont {Okazaki}\ \emph {et~al.}(2011)\citenamefont
  {Okazaki}, \citenamefont {Shibauchi}, \citenamefont {Shi}, \citenamefont
  {Haga}, \citenamefont {Matsuda}, \citenamefont {Yamamoto}, \citenamefont
  {Onuki}, \citenamefont {Ikeda},\ and\ \citenamefont {Matusuda}}]{okazaki:11}%
  \BibitemOpen
  \bibfield  {author} {\bibinfo {author} {\bibfnamefont {R.}~\bibnamefont
  {Okazaki}}, \bibinfo {author} {\bibfnamefont {T.}~\bibnamefont {Shibauchi}},
  \bibinfo {author} {\bibfnamefont {H.~J.}\ \bibnamefont {Shi}}, \bibinfo
  {author} {\bibfnamefont {Y.}~\bibnamefont {Haga}}, \bibinfo {author}
  {\bibfnamefont {T.}~\bibnamefont {Matsuda}}, \bibinfo {author} {\bibfnamefont
  {E.}~\bibnamefont {Yamamoto}}, \bibinfo {author} {\bibfnamefont
  {Y.}~\bibnamefont {Onuki}}, \bibinfo {author} {\bibfnamefont
  {H.}~\bibnamefont {Ikeda}}, \ and\ \bibinfo {author} {\bibfnamefont
  {Y.}~\bibnamefont {Matusuda}},\ }\href@noop {} {\bibfield  {journal}
  {\bibinfo  {journal} {Science}\ }\textbf {\bibinfo {volume} {331}},\ \bibinfo
  {pages} {439} (\bibinfo {year} {2011})}\BibitemShut {NoStop}%
\bibitem [{\citenamefont {Tonegawa}\ \emph {et~al.}(2012)\citenamefont
  {Tonegawa}, \citenamefont {Hashimoto}, \citenamefont {Ikada}, \citenamefont
  {Lin}, \citenamefont {Shishido}, \citenamefont {Haga}, \citenamefont
  {Matsuda}, \citenamefont {Yamamoto}, \citenamefont {Onuki}, \citenamefont
  {Ikeda}, \citenamefont {Matsuda},\ and\ \citenamefont
  {Shibauchi}}]{tonegawa:12}%
  \BibitemOpen
  \bibfield  {author} {\bibinfo {author} {\bibfnamefont {S.}~\bibnamefont
  {Tonegawa}}, \bibinfo {author} {\bibfnamefont {K.}~\bibnamefont {Hashimoto}},
  \bibinfo {author} {\bibfnamefont {K.}~\bibnamefont {Ikada}}, \bibinfo
  {author} {\bibfnamefont {Y.-H.}\ \bibnamefont {Lin}}, \bibinfo {author}
  {\bibfnamefont {H.}~\bibnamefont {Shishido}}, \bibinfo {author}
  {\bibfnamefont {Y.}~\bibnamefont {Haga}}, \bibinfo {author} {\bibfnamefont
  {T.~D.}\ \bibnamefont {Matsuda}}, \bibinfo {author} {\bibfnamefont
  {E.}~\bibnamefont {Yamamoto}}, \bibinfo {author} {\bibfnamefont
  {Y.}~\bibnamefont {Onuki}}, \bibinfo {author} {\bibfnamefont
  {H.}~\bibnamefont {Ikeda}}, \bibinfo {author} {\bibfnamefont
  {Y.}~\bibnamefont {Matsuda}}, \ and\ \bibinfo {author} {\bibfnamefont
  {T.}~\bibnamefont {Shibauchi}},\ }\href@noop {} {\bibfield  {journal}
  {\bibinfo  {journal} {Phys. Rev. Lett.}\ }\textbf {\bibinfo {volume} {109}},\
  \bibinfo {pages} {036401} (\bibinfo {year} {2012})}\BibitemShut {NoStop}%
\bibitem [{\citenamefont {Tonegawa}\ \emph {et~al.}(2014)\citenamefont
  {Tonegawa}, \citenamefont {Kasahara}, \citenamefont {Fukuda}, \citenamefont
  {Sugimoto}, \citenamefont {Yasuda}, \citenamefont {Tsuruhara}, \citenamefont
  {Watanabe}, \citenamefont {Mizukami}, \citenamefont {Haga}, \citenamefont
  {Matsuda}, \citenamefont {Yamamoto}, \citenamefont {Onuki}, \citenamefont
  {Ikeda}, \citenamefont {Matsuda},\ and\ \citenamefont
  {Shibauchi}}]{tonegawa:14}%
  \BibitemOpen
  \bibfield  {author} {\bibinfo {author} {\bibfnamefont {S.}~\bibnamefont
  {Tonegawa}}, \bibinfo {author} {\bibfnamefont {S.}~\bibnamefont {Kasahara}},
  \bibinfo {author} {\bibfnamefont {T.}~\bibnamefont {Fukuda}}, \bibinfo
  {author} {\bibfnamefont {K.}~\bibnamefont {Sugimoto}}, \bibinfo {author}
  {\bibfnamefont {N.}~\bibnamefont {Yasuda}}, \bibinfo {author} {\bibfnamefont
  {Y.}~\bibnamefont {Tsuruhara}}, \bibinfo {author} {\bibfnamefont
  {D.}~\bibnamefont {Watanabe}}, \bibinfo {author} {\bibfnamefont
  {Y.}~\bibnamefont {Mizukami}}, \bibinfo {author} {\bibfnamefont
  {Y.}~\bibnamefont {Haga}}, \bibinfo {author} {\bibfnamefont {T.~D.}\
  \bibnamefont {Matsuda}}, \bibinfo {author} {\bibfnamefont {E.}~\bibnamefont
  {Yamamoto}}, \bibinfo {author} {\bibfnamefont {Y.}~\bibnamefont {Onuki}},
  \bibinfo {author} {\bibfnamefont {H.}~\bibnamefont {Ikeda}}, \bibinfo
  {author} {\bibfnamefont {Y.}~\bibnamefont {Matsuda}}, \ and\ \bibinfo
  {author} {\bibfnamefont {T.}~\bibnamefont {Shibauchi}},\ }\href@noop {}
  {\bibfield  {journal} {\bibinfo  {journal} {Nature Communications}\ }\textbf
  {\bibinfo {volume} {5}},\ \bibinfo {pages} {4188} (\bibinfo {year}
  {2014})}\BibitemShut {NoStop}%
\bibitem [{\citenamefont {S.Takagi}\ \emph {et~al.}(2012)\citenamefont
  {S.Takagi}, \citenamefont {Ishihara}, \citenamefont {Yokoyama},\ and\
  \citenamefont {Amitsuka}}]{takagi:12}%
  \BibitemOpen
  \bibfield  {author} {\bibinfo {author} {\bibnamefont {S.Takagi}}, \bibinfo
  {author} {\bibfnamefont {S.}~\bibnamefont {Ishihara}}, \bibinfo {author}
  {\bibfnamefont {M.}~\bibnamefont {Yokoyama}}, \ and\ \bibinfo {author}
  {\bibfnamefont {H.}~\bibnamefont {Amitsuka}},\ }\href@noop {} {\bibfield
  {journal} {\bibinfo  {journal} {J. Phys. Soc. Jpn.}\ }\textbf {\bibinfo
  {volume} {81}},\ \bibinfo {pages} {114710} (\bibinfo {year}
  {2012})}\BibitemShut {NoStop}%
\bibitem [{\citenamefont {Kawasaki}\ \emph {et~al.}(2014)\citenamefont
  {Kawasaki}, \citenamefont {Watanabe}, \citenamefont {Hillier},\ and\
  \citenamefont {Aoki}}]{kawasaki:14}%
  \BibitemOpen
  \bibfield  {author} {\bibinfo {author} {\bibfnamefont {I.}~\bibnamefont
  {Kawasaki}}, \bibinfo {author} {\bibfnamefont {I.}~\bibnamefont {Watanabe}},
  \bibinfo {author} {\bibfnamefont {A.}~\bibnamefont {Hillier}}, \ and\
  \bibinfo {author} {\bibfnamefont {D.}~\bibnamefont {Aoki}},\ }\href@noop {}
  {\bibfield  {journal} {\bibinfo  {journal} {J. Phys. Soc. Jpn.}\ }\textbf
  {\bibinfo {volume} {83}},\ \bibinfo {pages} {094720} (\bibinfo {year}
  {2014})}\BibitemShut {NoStop}%
\bibitem [{\citenamefont {Kusunose}\ and\ \citenamefont
  {Harima}(2011)}]{kusunose:11}%
  \BibitemOpen
  \bibfield  {author} {\bibinfo {author} {\bibfnamefont {H.}~\bibnamefont
  {Kusunose}}\ and\ \bibinfo {author} {\bibfnamefont {H.}~\bibnamefont
  {Harima}},\ }\href@noop {} {\bibfield  {journal} {\bibinfo  {journal} {J.
  Phys.Soc. Jpn.}\ }\textbf {\bibinfo {volume} {80}},\ \bibinfo {pages}
  {084702} (\bibinfo {year} {2011})}\BibitemShut {NoStop}%
\bibitem [{\citenamefont {Capriotti}\ \emph {et~al.}(2003)\citenamefont
  {Capriotti}, \citenamefont {Scalapino},\ and\ \citenamefont
  {Sedgewick}}]{capriotti:03}%
  \BibitemOpen
  \bibfield  {author} {\bibinfo {author} {\bibfnamefont {L.}~\bibnamefont
  {Capriotti}}, \bibinfo {author} {\bibfnamefont {D.~J.}\ \bibnamefont
  {Scalapino}}, \ and\ \bibinfo {author} {\bibfnamefont {R.~D.}\ \bibnamefont
  {Sedgewick}},\ }\href@noop {} {\bibfield  {journal} {\bibinfo  {journal}
  {Phys. Rev. B}\ }\textbf {\bibinfo {volume} {68}},\ \bibinfo {pages} {014508}
  (\bibinfo {year} {2003})}\BibitemShut {NoStop}%
\bibitem [{\citenamefont {Akbari}\ and\ \citenamefont
  {Thalmeier}(2013{\natexlab{a}})}]{akbari:13}%
  \BibitemOpen
  \bibfield  {author} {\bibinfo {author} {\bibfnamefont {A.}~\bibnamefont
  {Akbari}}\ and\ \bibinfo {author} {\bibfnamefont {P.}~\bibnamefont
  {Thalmeier}},\ }\href@noop {} {\bibfield  {journal} {\bibinfo  {journal}
  {Eur. Phys. J. B}\ }\textbf {\bibinfo {volume} {86}},\ \bibinfo {pages} {495}
  (\bibinfo {year} {2013}{\natexlab{a}})}\BibitemShut {NoStop}%
\bibitem [{\citenamefont {Akbari}\ and\ \citenamefont
  {Thalmeier}(2013{\natexlab{b}})}]{akbari:13b}%
  \BibitemOpen
  \bibfield  {author} {\bibinfo {author} {\bibfnamefont {A.}~\bibnamefont
  {Akbari}}\ and\ \bibinfo {author} {\bibfnamefont {P.}~\bibnamefont
  {Thalmeier}},\ }\href {http://stacks.iop.org/0295-5075/102/i=5/a=57008}
  {\bibfield  {journal} {\bibinfo  {journal} {EPL (Europhysics Letters)}\
  }\textbf {\bibinfo {volume} {102}},\ \bibinfo {pages} {57008} (\bibinfo
  {year} {2013}{\natexlab{b}})}\BibitemShut {NoStop}%
\bibitem [{\citenamefont {Amitsuka}\ \emph {et~al.}(2007)\citenamefont
  {Amitsuka}, \citenamefont {Matsuda}, \citenamefont {Kawasaki}, \citenamefont
  {Tenya}, \citenamefont {Yokoyama}, \citenamefont {Sekine}, \citenamefont
  {Tateiwa}, \citenamefont {Kobayashi}, \citenamefont {Kawarazaki},\ and\
  \citenamefont {Yoshizawa}}]{amitsuka:07}%
  \BibitemOpen
  \bibfield  {author} {\bibinfo {author} {\bibfnamefont {H.}~\bibnamefont
  {Amitsuka}}, \bibinfo {author} {\bibfnamefont {K.}~\bibnamefont {Matsuda}},
  \bibinfo {author} {\bibfnamefont {I.}~\bibnamefont {Kawasaki}}, \bibinfo
  {author} {\bibfnamefont {K.}~\bibnamefont {Tenya}}, \bibinfo {author}
  {\bibfnamefont {M.}~\bibnamefont {Yokoyama}}, \bibinfo {author}
  {\bibfnamefont {C.}~\bibnamefont {Sekine}}, \bibinfo {author} {\bibfnamefont
  {N.}~\bibnamefont {Tateiwa}}, \bibinfo {author} {\bibfnamefont {T.~C.}\
  \bibnamefont {Kobayashi}}, \bibinfo {author} {\bibfnamefont {S.}~\bibnamefont
  {Kawarazaki}}, \ and\ \bibinfo {author} {\bibfnamefont {H.}~\bibnamefont
  {Yoshizawa}},\ }\href@noop {} {\bibfield  {journal} {\bibinfo  {journal} {J.
  Magn. Magn. Mater.}\ }\textbf {\bibinfo {volume} {310}},\ \bibinfo {pages}
  {214} (\bibinfo {year} {2007})}\BibitemShut {NoStop}%
\bibitem [{\citenamefont {Kasahara}\ \emph {et~al.}(2009)\citenamefont
  {Kasahara}, \citenamefont {Shishido}, \citenamefont {Shibauchi},
  \citenamefont {Haga}, \citenamefont {Matsuda}, \citenamefont {Onuki},\ and\
  \citenamefont {Matsuda}}]{kasahara:09}%
  \BibitemOpen
  \bibfield  {author} {\bibinfo {author} {\bibfnamefont {Y.}~\bibnamefont
  {Kasahara}}, \bibinfo {author} {\bibfnamefont {H.}~\bibnamefont {Shishido}},
  \bibinfo {author} {\bibfnamefont {T.}~\bibnamefont {Shibauchi}}, \bibinfo
  {author} {\bibfnamefont {Y.}~\bibnamefont {Haga}}, \bibinfo {author}
  {\bibfnamefont {T.~D.}\ \bibnamefont {Matsuda}}, \bibinfo {author}
  {\bibfnamefont {Y.}~\bibnamefont {Onuki}}, \ and\ \bibinfo {author}
  {\bibfnamefont {Y.}~\bibnamefont {Matsuda}},\ }\href@noop {} {\bibfield
  {journal} {\bibinfo  {journal} {New Journal of Physics}\ }\textbf {\bibinfo
  {volume} {11}},\ \bibinfo {pages} {055061} (\bibinfo {year}
  {2009})}\BibitemShut {NoStop}%
\bibitem [{\citenamefont {Yano}\ \emph {et~al.}(2008)\citenamefont {Yano},
  \citenamefont {Sakakibara}, \citenamefont {Tayama}, \citenamefont {Yokoyama},
  \citenamefont {Amitsuka}, \citenamefont {Homma}, \citenamefont {Miranovic},
  \citenamefont {Ichioka}, \citenamefont {Tsutsumi},\ and\ \citenamefont
  {Machida}}]{yano:08}%
  \BibitemOpen
  \bibfield  {author} {\bibinfo {author} {\bibfnamefont {K.}~\bibnamefont
  {Yano}}, \bibinfo {author} {\bibfnamefont {T.}~\bibnamefont {Sakakibara}},
  \bibinfo {author} {\bibfnamefont {T.}~\bibnamefont {Tayama}}, \bibinfo
  {author} {\bibfnamefont {M.}~\bibnamefont {Yokoyama}}, \bibinfo {author}
  {\bibfnamefont {H.}~\bibnamefont {Amitsuka}}, \bibinfo {author}
  {\bibfnamefont {Y.}~\bibnamefont {Homma}}, \bibinfo {author} {\bibfnamefont
  {P.}~\bibnamefont {Miranovic}}, \bibinfo {author} {\bibfnamefont
  {M.}~\bibnamefont {Ichioka}}, \bibinfo {author} {\bibfnamefont
  {Y.}~\bibnamefont {Tsutsumi}}, \ and\ \bibinfo {author} {\bibfnamefont
  {K.}~\bibnamefont {Machida}},\ }\href@noop {} {\bibfield  {journal} {\bibinfo
   {journal} {Phys. Rev. Lett.}\ }\textbf {\bibinfo {volume} {100}},\ \bibinfo
  {pages} {017004} (\bibinfo {year} {2008})}\BibitemShut {NoStop}%
\bibitem [{\citenamefont {Park}\ \emph {et~al.}(2012)\citenamefont {Park},
  \citenamefont {Tobash}, \citenamefont {Ronning}, \citenamefont {Bauer},
  \citenamefont {Sarrao}, \citenamefont {Thompson},\ and\ \citenamefont
  {Greene}}]{park:12}%
  \BibitemOpen
  \bibfield  {author} {\bibinfo {author} {\bibfnamefont {W.~K.}\ \bibnamefont
  {Park}}, \bibinfo {author} {\bibfnamefont {P.~H.}\ \bibnamefont {Tobash}},
  \bibinfo {author} {\bibfnamefont {F.}~\bibnamefont {Ronning}}, \bibinfo
  {author} {\bibfnamefont {E.~D.}\ \bibnamefont {Bauer}}, \bibinfo {author}
  {\bibfnamefont {J.~L.}\ \bibnamefont {Sarrao}}, \bibinfo {author}
  {\bibfnamefont {J.~D.}\ \bibnamefont {Thompson}}, \ and\ \bibinfo {author}
  {\bibfnamefont {L.~H.}\ \bibnamefont {Greene}},\ }\href@noop {} {\bibfield
  {journal} {\bibinfo  {journal} {Phys. Rev. Lett.}\ }\textbf {\bibinfo
  {volume} {108}},\ \bibinfo {pages} {246403} (\bibinfo {year}
  {2012})}\BibitemShut {NoStop}%
\bibitem [{\citenamefont {Aynajian}\ \emph {et~al.}(2010)\citenamefont
  {Aynajian}, \citenamefont {da~Silva~Neto}, \citenamefont {Parker},
  \citenamefont {Huang}, \citenamefont {Pasupathy}, \citenamefont {Mydosh},\
  and\ \citenamefont {Yazdani}}]{aynajian:10}%
  \BibitemOpen
  \bibfield  {author} {\bibinfo {author} {\bibfnamefont {P.}~\bibnamefont
  {Aynajian}}, \bibinfo {author} {\bibfnamefont {E.~H.}\ \bibnamefont
  {da~Silva~Neto}}, \bibinfo {author} {\bibfnamefont {C.~V.}\ \bibnamefont
  {Parker}}, \bibinfo {author} {\bibfnamefont {Y.}~\bibnamefont {Huang}},
  \bibinfo {author} {\bibfnamefont {A.}~\bibnamefont {Pasupathy}}, \bibinfo
  {author} {\bibfnamefont {J.}~\bibnamefont {Mydosh}}, \ and\ \bibinfo {author}
  {\bibfnamefont {A.}~\bibnamefont {Yazdani}},\ }\href@noop {} {\bibfield
  {journal} {\bibinfo  {journal} {Proc. Nat. Acad. Sci.}\ }\textbf {\bibinfo
  {volume} {107}},\ \bibinfo {pages} {10383} (\bibinfo {year}
  {2010})}\BibitemShut {NoStop}%
\bibitem [{\citenamefont {Okazaki}\ \emph {et~al.}(2010)\citenamefont
  {Okazaki}, \citenamefont {Shimozawa}, \citenamefont {Shishido}, \citenamefont
  {Konczykowski}, \citenamefont {Haga}, \citenamefont {Matsuda}, \citenamefont
  {Yamamoto}, \citenamefont {Onuki}, \citenamefont {Yanase}, \citenamefont
  {Shibauchi},\ and\ \citenamefont {Matsuda}}]{okazaki:10}%
  \BibitemOpen
  \bibfield  {author} {\bibinfo {author} {\bibfnamefont {R.}~\bibnamefont
  {Okazaki}}, \bibinfo {author} {\bibfnamefont {M.}~\bibnamefont {Shimozawa}},
  \bibinfo {author} {\bibfnamefont {H.}~\bibnamefont {Shishido}}, \bibinfo
  {author} {\bibfnamefont {M.}~\bibnamefont {Konczykowski}}, \bibinfo {author}
  {\bibfnamefont {Y.}~\bibnamefont {Haga}}, \bibinfo {author} {\bibfnamefont
  {T.~D.}\ \bibnamefont {Matsuda}}, \bibinfo {author} {\bibfnamefont
  {E.}~\bibnamefont {Yamamoto}}, \bibinfo {author} {\bibfnamefont
  {Y.}~\bibnamefont {Onuki}}, \bibinfo {author} {\bibfnamefont
  {Y.}~\bibnamefont {Yanase}}, \bibinfo {author} {\bibfnamefont
  {T.}~\bibnamefont {Shibauchi}}, \ and\ \bibinfo {author} {\bibfnamefont
  {Y.}~\bibnamefont {Matsuda}},\ }\href@noop {} {\bibfield  {journal} {\bibinfo
   {journal} {J. Phys.Soc. Jpn.}\ }\textbf {\bibinfo {volume} {79}},\ \bibinfo
  {pages} {084705} (\bibinfo {year} {2010})}\BibitemShut {NoStop}%
\bibitem [{\citenamefont {Bel}\ \emph {et~al.}(2004)\citenamefont {Bel},
  \citenamefont {Jin}, \citenamefont {Behnia}, \citenamefont {Flouquet},\ and\
  \citenamefont {Lejay}}]{bel:04}%
  \BibitemOpen
  \bibfield  {author} {\bibinfo {author} {\bibfnamefont {R.}~\bibnamefont
  {Bel}}, \bibinfo {author} {\bibfnamefont {H.}~\bibnamefont {Jin}}, \bibinfo
  {author} {\bibfnamefont {K.}~\bibnamefont {Behnia}}, \bibinfo {author}
  {\bibfnamefont {J.}~\bibnamefont {Flouquet}}, \ and\ \bibinfo {author}
  {\bibfnamefont {P.}~\bibnamefont {Lejay}},\ }\href@noop {} {\bibfield
  {journal} {\bibinfo  {journal} {Phys. Rev. B}\ }\textbf {\bibinfo {volume}
  {70}},\ \bibinfo {pages} {220501} (\bibinfo {year} {2004})}\BibitemShut
  {NoStop}%
\bibitem [{\citenamefont {Behnia}\ \emph {et~al.}(2005)\citenamefont {Behnia},
  \citenamefont {Bel}, \citenamefont {Kasahara}, \citenamefont {Nakajima},
  \citenamefont {Jin}, \citenamefont {Aubin}, \citenamefont {Izawa},
  \citenamefont {Matsuda}, \citenamefont {Flouquet}, \citenamefont {Haga},
  \citenamefont {Onuki},\ and\ \citenamefont {Lejay}}]{behnia:05}%
  \BibitemOpen
  \bibfield  {author} {\bibinfo {author} {\bibfnamefont {K.}~\bibnamefont
  {Behnia}}, \bibinfo {author} {\bibfnamefont {R.}~\bibnamefont {Bel}},
  \bibinfo {author} {\bibfnamefont {Y.}~\bibnamefont {Kasahara}}, \bibinfo
  {author} {\bibfnamefont {Y.}~\bibnamefont {Nakajima}}, \bibinfo {author}
  {\bibfnamefont {H.}~\bibnamefont {Jin}}, \bibinfo {author} {\bibfnamefont
  {H.}~\bibnamefont {Aubin}}, \bibinfo {author} {\bibfnamefont
  {K.}~\bibnamefont {Izawa}}, \bibinfo {author} {\bibfnamefont
  {Y.}~\bibnamefont {Matsuda}}, \bibinfo {author} {\bibfnamefont
  {J.}~\bibnamefont {Flouquet}}, \bibinfo {author} {\bibfnamefont
  {Y.}~\bibnamefont {Haga}}, \bibinfo {author} {\bibfnamefont {Y.}~\bibnamefont
  {Onuki}}, \ and\ \bibinfo {author} {\bibfnamefont {P.}~\bibnamefont
  {Lejay}},\ }\href@noop {} {\bibfield  {journal} {\bibinfo  {journal} {Phys.
  Rev. Lett.}\ }\textbf {\bibinfo {volume} {94}},\ \bibinfo {pages} {156405}
  (\bibinfo {year} {2005})}\BibitemShut {NoStop}%
\end{thebibliography}%

\end{document}